\pdfoutput=1
\documentclass[runningheads]{llncs}
\usepackage{comment}
\usepackage{graphicx}
\usepackage{lmodern} 

\usepackage{listings, boogie}
\usepackage{color}
\usepackage{xspace}
\usepackage{cite}
\usepackage{amsmath, mathtools}
\usepackage{graphbox,graphicx}
\definecolor{light-gray}{gray}{0.9}
\definecolor{darkgreen}{rgb}{0.0,0.7,0.0}
\definecolor{NewCmdColor}{rgb}{0.0, 0.0, 1.0}
\definecolor{CertLinkColor}{rgb}{0.0,0.3,0.55}
\definecolor{FinalCertLinkColor}{rgb}{0.0,0.7,0.0}

\usepackage{tikz}
\usetikzlibrary{arrows, positioning}

\newcommand{\wrt}{{{w.r.t.\@}}}

\newcommand{\eg}{{{e.g.\@}}}
\newcommand{\ie}{{{i.e.\@}}}
\newcommand{\cf}{{\it{cf.\@}}}

\newcommand\secref[1]{Sec.~\ref{#1}} 

\newcommand\tabref[1]{Tab.~\ref{#1}}
\newcommand\figref[1]{Fig.~\ref{#1}}

\def\trref{\ifx\istr\undefined{ of the TR~\cite{TR}}\fi}
\def\trrefnocite{\ifx\istr\undefined{ of the TR}\fi}
\newcommand\appreftr[1]{App. #1\trref{}}

\usepackage[normalem]{ulem}

\def\nocolour{ }

\newcommand{\soutifcolour}[1]{\ifdefined\nocolour{}\else{\sout{#1}}\fi}

\newcommand{\as}[1]{\ifdefined\nocolour{{#1}}\else{\color{orange!70!black}{#1}}\fi}
\newcommand{\asfootnote}[1]{\ifdefined\nocolour{}\else\as{\footnote{\as{ALEX: #1}}}\fi}
\newcommand{\asout}[1]{\as{\soutifcolour{#1}}}

\newcommand{\gaurav}[1]{\ifdefined\nocolour{#1}\else{\color{darkgreen}{#1}}\fi}

\newcommand{\gout}[1]{\gaurav{{\soutifcolour{#1}}}}

\newcommand\todo[1]{\ifdefined\nocolour{}\else{\textcolor{red}{TODO: #1}}\fi}

\newcommand\inl[1]{\textsf{inl}(#1)}
\newcommand\inr[1]{\textsf{inr}(#1)}

\newcommand\globversion[1]{\ensuremath{\textsf{ver}_{\mathcal{G}}(#1)}}

\newcommand\wpboogie[1]{\ensuremath{wp(#1,\btrue{})}}

\definecolor{darkred}{rgb}{0.55, 0.0, 0.0}

\usepackage{prooftree}

\makeatletter

\newskip \point

\def \premisespacing{\quad}

\def \RulePremisesNewlineMore[#1]#2.#3#4{\@ifnextchar\bgroup{\RulePremisesNewlineMore[#1]{#2}.{#3\premisespacing#4}}{\@ifnextchar.{\RulePremisesNewline[#1]{{\begin{array}{c}#2\\#3\premisespacing#4\end{array}}}}{\RuleMultiPremise[#1]{{\begin{array}{c}#2\\#3\end{array}}}{#4}}}}

\def \RulePremisesNewline[#1]#2.#3{\@ifnextchar\bgroup{\RulePremisesNewlineMore[#1]{#2}.{#3}}{\@ifnextchar.{\RulePremisesNewline[#1]{{\begin{array}{c}#2\\#3\end{array}}}}{\RuleMultiPremise[#1]{#2}{#3}}}}

\def \RuleMultiPremise[#1]#2#3{\@ifnextchar\bgroup{\RuleMultiPremise[#1]{#2\premisespacing#3}}{\@ifnextchar.{\RulePremisesNewline[#1]{#2\premisespacing#3}}{\prooftree #2\justifies#3 \using{#1}\endprooftree}}}

\def \RuleWithName[#1]#2{\@ifnextchar\bgroup {\RuleMultiPremise[#1]{#2}}{\@ifnextchar.{\RulePremisesNewline[#1]{#2}}{\prooftree \justifies #2 \using{#1} \endprooftree}}}

\def \RuleWithInfo[#1]{\@ifnextchar[{\RuleWithNameAndCondition[#1]}{\RuleWithName[(#1)]}}

\def \RuleWithNameAndCondition[#1][#2]{\RuleWithName[(#1)^{#2}]}

\def \Inf{\proofrulebaseline=2ex \abovedisplayskip12\point\belowdisplayskip12\point \abovedisplayshortskip8\point\belowdisplayshortskip8\point \@ifnextchar[{\RuleWithInfo}{\RuleWithName[ ]}}

\makeatother

\makeatletter
\def\operator#1{\@ifnextchar\bgroup {\operatorarg{\ensuremath{#1}}}{\ensuremath{#1}}}
\def\operatorarg#1#2{{#1}{\ensuremath{(#2)}}}
\def\spoperator#1#2{\@ifnextchar\bgroup{\spoperatorarg{\ensuremath{#1}}{\ensuremath{#2}}}{\ensuremath{#1}}}
\def\spoperatorarg#1#2#3{\ensuremath{#1#2#3}}

\def\fixedoperator#1{\@ifnextchar\bgroup {\fixedoperatorarg{#1}}{\ensuremath{#1}}}
\def\fixedoperatorarg#1#2{\fixedoperatorparse{#1}#2~}
\def\fixedoperatorparse#1#2,#3~{\ensuremath{{#2}{.}{#1}{(#3)}}}
\makeatother

\newskip \point \point =1pt
\setbox134=\hbox{\leavevmode\raise0\point\hbox{${\langle}\kern-2.5\point{\langle}$}}
\setbox135=\hbox{\raise0\point\hbox{${ \rangle}\kern-2.5\point{ \rangle}$}}

\setbox136=\hbox{\leavevmode\raise0\point\hbox{${\lfloor}\kern-3.0\point{\lfloor}$}}
\setbox137=\hbox{\raise0\point\hbox{${ \rfloor}\kern-3.0\point{ \rfloor}$}}

\setbox138=\hbox{\leavevmode\raise0\point\hbox{${[}\kern-1.5\point{[}$}}
\setbox139=\hbox{\raise0\point\hbox{${]}\kern-1.5\point{]}$}}

\newcommand{\typs}{\ensuremath{\mathcal{T}}} 

\newcommand{\modX}{\ensuremath{X_H}} 
\newcommand{\block}[1]{\mathit{cs}_{#1}} 
\newcommand{\srcBlock}{\block{S}}
\newcommand{\trgBlock}{\block{T}}

\newcommand{\varRel}{\ensuremath{\mathcal{V}_R}}
\newcommand{\srcCmds}{\ensuremath{\mathit{cs}}}
\newcommand{\trgCmds}{\ensuremath{\mathit{cs}'}}

\usepackage[T1]{fontenc}
\usepackage[scaled=0.81]{beramono}

\def\istr{} 

\begin{document}

\title{Formally Validating\\a Practical Verification Condition Generator (extended version)}
%
%
\author{ Gaurav Parthasarathy \inst{1} \and Peter M\"uller \inst{1} \and  Alexander J. Summers \inst{2}}
\titlerunning{Formally Validating a Practical Verification Condition Generator}
\authorrunning{G. Parthasarathy et al.}
%
\institute{Department of Computer Science, ETH Zurich, Switzerland\\
\email{\{gaurav.parthasarathy,peter.mueller\}@inf.ethz.ch} \and
University of British Columbia\\
\email{alex.summers@ubc.ca}
}
\maketitle              
\begin{abstract}
A program verifier produces reliable results only if both the \emph{logic} used to justify the program's correctness is sound, and the \emph{implementation} of the program verifier is itself correct.
Whereas it is common to formally prove soundness of the logic, the implementation of a verifier typically remains unverified. Bugs in verifier implementations may compromise the trustworthiness of successful verification results. Since program verifiers used in practice are complex, evolving software systems, it is generally not feasible to formally verify their implementation.

In this paper, we present an alternative approach: we \emph{validate successful runs} of the widely-used Boogie verifier by producing a \emph{certificate} which proves correctness of the obtained verification result. Boogie performs a complex series of program translations before ultimately generating a verification condition whose validity should imply the correctness of the input program. We show how to certify three of Boogie's core transformation phases: the elimination of cyclic control flow paths, the (SSA-like) replacement of assignments by assumptions using fresh variables (passification), and the final generation of verification conditions. Similar translations are employed by other verifiers. Our implementation produces certificates in Isabelle, based on a novel formalisation of the Boogie language.

\end{abstract}

\section{Introduction}\label{sec:intro}
Program verifiers are tools which attempt to prove the correctness of an implementation with respect to its specification. A successful verification attempt is, however, only meaningful if both the \emph{logic} used to justify the program's correctness is sound, and the \emph{implementation} of the program verifier is itself correct.
It is common to formally prove soundness of the logic, but the implementations of program verifiers typically remain unverified. As is standard for complex software systems, bugs in verifier implementations can and do arise, potentially raising doubts as to the trustworthiness of successful verification results.

One way to close this gap is to prove a verifier's implementation correct. However, such a \emph{once-and-for-all} approach faces serious challenges. Verifying an existing implementation bottom-up is not practically feasible because such implementations tend to be large and complex (for instance, the Boogie verifier~\cite{boogie} consists of over 30K lines of imperative C\# code), use a variety of libraries, and are typically written in efficient mainstream programming languages which themselves lack a formalisation. Alternatively, one could develop a verifier that is correct by construction. However, this approach requires the verifier to be (re-)implemented in an interactive theorem prover (ITP) such as Coq~\cite{Coq} or Isabelle~\cite{Isabelle}. This precludes the free choice of implementation language and paradigm, exploitation of concurrency, and possibility of tight integration with standard compilers and IDEs, which is often desirable for program verifiers~\cite{AstrauskasMuellerPoliSummers19b,BarnettEA10,framac15,vcc}.
Both verification approaches substantially impede software maintenance, which is problematic since verifiers are often rapidly-evolving software projects (for instance, the Boogie repository~\cite{boogierepo} contains more than 5000 commits). 
\gout{It is, thus, not surprising that once-and-for-all verification of program verifiers has been restricted to idealised implementations, omitting for instance challenging optimisations;
the gap between those implementations and the tools used in practice remains significant.}

To address these challenges, in this work we employ a different approach. Instead of verifying the implementation once and for all, we \emph{validate specific runs} of the verifier by \gaurav{automatically} producing a \emph{certificate} which proves the correctness of the obtained verification result. Our certificate generation formally relates the input and output of the verifier, but does so largely independently of its implementation, which can freely employ complex languages,
algorithms, or optimisations. Our certificates are formal proofs in Isabelle, and so checkable by an independent trusted tool; their guarantees for a certified run of the verifier are as strong as those provided by a (hypothetical) verified verifier.

We apply our novel verifier validation approach to the widely-used Boogie verifier, which verifies programs written in the intermediate verification language Boogie. 
The Boogie verifier is a \emph{verification condition generator}: it verifies programs by generating a verification condition (VC), whose validity is then discharged by an SMT solver. Certifying a verifier run requires proving that validity of the VC implies the correctness of the input program. Certification of the validity-checking of the VC is an orthogonal concern; our results can be combined with work in that area~\cite{BohmeW10,EkiciMTKKRB17,verit-hol-2019} to obtain end-to-end guarantees.

Like many automatic verifiers, Boogie is a \emph{translational verifier}: it performs a sequence of substantial Boogie-to-Boogie translations (\emph{phases}), simplifying the task and output of the final efficient VC computation~\cite{FlanaganSaxe2001,BarnettLeino2005}. The key challenges in certifying runs of the Boogie tool are to certify each of these phases, including final VC generation. In particular, we present novel techniques for making the following three key phases (and many smaller ones) of Boogie's tool chain certifying:

\begin{enumerate}
\item
The elimination of loops (more precisely, cycles in the CFG) by reducing the correctness of loops to checking loop invariants \emph{(CFG-to-DAG phase)}

\item
The replacement of assignments by (SSA-style) introduction of fresh variables and suitable \assumeNoArg{} statements \emph{(passification phase)}

\item
The final generation of the VC, which includes the erasure and logical encoding of Boogie's polymorphic type system~\cite{LeinoR10} \emph{(VC phase)}.
\end{enumerate}

The certification of such verifier phases is related to existing work on compiler verification~\cite{LeroyCompCert} and validation~\cite{Tristan2008,Tristan2009,compcertssa}. However, the translations and the certified property we tackle here are fundamentally different from those in compilers. Compilers typically require that each execution of the target program corresponds to an execution of the source program. In contrast, the encoding of a program in a translational verifier typically has intentionally more executions (for instance, allows more non-determinism). Moreover, translational verifiers need to handle features not present in standard programming languages such as \assumeNoArg{} statements and background theories. Prior work on validating such verifier phases has been limited in the supported language and extent of the formal guarantee; we discuss comparisons in detail in \secref{sec:related_work}.

\subsubsection{Contributions.} Our paper makes the following technical contributions.

\begin{enumerate}
\item The first formal semantics for a significant subset of Boogie (including axioms, polymorphism, type constructors), mechanised in Isabelle.

\item A validation technique for two core program-to-program translations occurring in verifiers (CFG-to-DAG and passification).

\item A validation technique for the VC phase, handling polymorphism erasure and Boogie's type system encoding~\cite{Leino10}, for which no prior formal proof exists.

\item A version of the Boogie implementation that produces certificates for a significant subset of Boogie.
\end{enumerate}

Making the Boogie verifier certifying is an important result, reducing the trusted code base for a wide variety of verification tools implemented via encodings into Boogie, \eg{} Dafny~\cite{Leino10}, VCC~\cite{vcc}, Corral~\cite{LalQL12}, and Viper~\cite{MuellerSchwerhoffSummers16}. Moreover, the technical approach we present here can in future be applied to the certification of the translations performed by these tools, and those based on comparable intermediate verification languages such as Frama-C~\cite{framac15} and Krakatoa~\cite{FilliatreM07} based on Why3~\cite{why3} and
Prusti~\cite{AstrauskasMuellerPoliSummers19b} and
VerCors~\cite{BlomDHO17} based on Viper~\cite{MuellerSchwerhoffSummers16}.

\paragraph{Outline.}
\secref{sec:overview} explains at a high-level, how our validation approach is structured for the different phases.
\secref{sec:boogie_lang} introduces a formal semantics for Boogie.
Secs.~\ref{sec:cfg_to_dag}, \ref{sec:passification} and \ref{sec:vc_phase} present our validation of the CFG-to-DAG, passification, and VC phases, respectively.
\secref{sec:evaluation} evaluates our certificate-producing version of Boogie.
\secref{sec:related_work} discusses related work. \secref{sec:conclusion} concludes. 
Further details are available in 
\ifdefined\istr the appendix. \else our accompanying technical report (hereafter, TR) \cite{TR}. \todo{update citation} \fi

\section{Approach}\label{sec:overview}
A Boogie program consists of a set of procedures, each with a specification and a procedure body in the form of a (reducible) control-flow-graph (CFG), whose blocks contain basic commands; we present the formal details in the next section. Boogie verifies each procedure modularly, desugaring procedure calls according to their specifications. Verification is implemented via a series of phases: program-to-program translations and a final computation of a VC to be checked by an SMT solver. Our goal is to formally certify (per run of Boogie) that validity of this VC implies the correctness of the original procedure.

To keep the complexity of certificates manageable, our technical approach is \emph{modular} in three dimensions: decomposing our formal goal per \emph{procedure} in the Boogie program, per \emph{phase} of the Boogie verification, and per \emph{block} in the CFG of each procedure. This modularity makes the full automation of our certification proofs in Isabelle practical. In the following, we give a high-level overview of this modular structure; the details are presented in subsequent sections.

\paragraph{Procedure decomposition.}
Boogie has no notion of a main program or an overall program execution. A Boogie program is correct if each of its procedures is individually correct (that is, the procedure body has no failing traces, as we make precise in the next section). Boogie computes a separate VC for each procedure, and we correspondingly validate the verification of each procedure separately.

\paragraph{Phase decomposition.}

\begin{figure}[t!]
    \centering
\begin{tikzpicture}[node distance=8em]

    \tikzset{vertex/.style = {shape=rectangle,draw,minimum size=1.5em, text width = 1.5 em}}
    \tikzset{nullvertex/.style = {shape=circle,draw,minimum size=1.5em}}
    \tikzset{edge/.style = {->,> = latex'}}
    
    \node[align=center, vertex](P1) {$G_1$};
    \node[align=center, vertex, right=of P1](P2) {$G_2$};
    \node[align=center, vertex, right=of P2](P3) {$G_3$};
    \node[align=center, vertex, right=of P3](P4) {VC};
    \node[align=center, below=3em of P4](P4Below) {};
    \node[align=center, below=3em of P1](P1Below) {};

    \path
    (P1) [->] edge node [above] {\small CFG-to-DAG} (P2)
    (P2) edge node [above] {\small Passification} (P3)
    (P3) edge node [above] {\small VC Phase} (P4)
    (P4) edge [densely dotted, bend left=30, color=CertLinkColor] node [below] {
        \small $\textsf{valid}(\text{VC}) \models \textsf{ver}(G_3)$} (P3)
    (P3) edge [densely dotted, bend left=30, color=CertLinkColor] node [below] {
        \small $\textsf{ver}(G_3) \models \textsf{ver}(G_2)$} (P2)
    (P2) edge [densely dotted, bend left=30, color=CertLinkColor] node [below] {
       \small $\textsf{ver}(G_2) \models \textsf{ver}(G_1)$} (P1)
    (P4) edge [dashed, color=FinalCertLinkColor, style={-}] node [below] {} (P4Below.center)
    (P4Below.center) edge [dashed, color=FinalCertLinkColor, style={-}] node [below] {\small $\textsf{valid}(\text{VC}) \models \textsf{ver}(G_1)$} (P1Below.center)
    (P1Below.center) edge [dashed, color=FinalCertLinkColor] node [below] {} (P1);

\end{tikzpicture}

\caption{Key phases of verification in Boogie and their certification. The solid edges show Boogie's transformations on a procedure body; each node $G_i$ represents a control-flow-graph. Our final certificate (\gout{green}\gaurav{dashed} edge) is constructed by formally linking the three phase certificates represented by the \gaurav{dotted}\gout{blue} edges.
Each of the three phase certificates also incorporate extra smaller transformations that we do not show here.
}
\label{fig:pipeline_overview}
\end{figure}
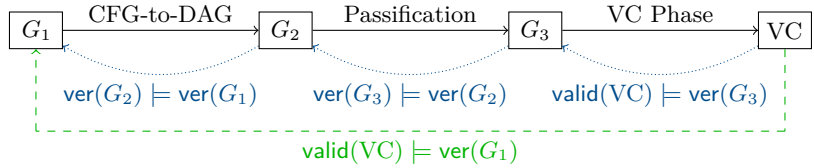

We break our overall validation efforts down into per-phase sub-problems. In this paper, we focus on the following three most substantial and technically-challenging of these sequential phases, illustrated in \figref{fig:pipeline_overview}.
(1)~The \emph{CFG-to-DAG phase} translates a (possibly-cyclic) CFG to an acyclic CFG (\cf{} \secref{sec:cfg_to_dag}). This phase substantially alters the CFG structure, cutting loops using annotated loop invariants to over-approximate their executions.
(2)~The \emph{passification phase} eliminates imperative updates by transforming the code into static single assignment (SSA) form and then replacing assignments with \emph{constraints} on variable versions (\cf{} \secref{sec:passification}). Both of these phases introduce extra non-determinism and \assumeNoArg{} statements (which, if implemented incorrectly could make verification unsound by masking errors in the program).
(3)~The final \emph{VC phase} translates the acyclic, passified CFG to a verification condition that, in addition to capturing the weakest precondition, encodes away Boogie's polymorphic type system~\cite{LeinoR10}.

We construct certificates for each of these key phases separately
(depicted by the \as{blue} \gaurav{dotted} lines in \figref{fig:pipeline_overview}).
For each phase, we certify that \emph{if} the target of the translation phase is correct (a correct Boogie program for the first two phases; a valid VC for the VC phase) then the source (program) of the phase is correct.  This modular approach lets us focus the proof strategy for each phase on its conceptually-relevant concerns, and provides robustness against \emph{changes} to the verifier since at most the certification of the changed phases may need adjustment. Logically, our per-phase certificates are finally glued together to guarantee the analogous end-to-end property for the entire pipeline, depicted \gout{in green}\gaurav{by the \as{green} dashed edge} in \figref{fig:pipeline_overview}.
\gaurav{For our certificates, we import the input and output programs (and VC) of each key phase from Boogie into Isabelle\as{;}\asout{ and} we do not reimplement \as{any of Boogie's} phases inside Isabelle.}

The certificates of the key phases also incorporate various smaller transformations between the key phases, such as peephole optimisation. Our work \as{also} \as{validates}\asout{covers} these small\as{er} transformations, but we focus \as{the presentation} on the key phases in \as{this} paper.
Boogie also performs several smaller translation steps \emph{prior} to the CFG-to-DAG phase. These include transforming ASTs to corresponding CFGs, optimisations such as dead variable elimination, and desugaring procedure calls using their specifications (via explicit \assertNoArg{}, \assumeNoArg{}, and \havocNoArg{} statements). Our approach applies analogously to these initial smaller phases, but our current implementation certifies only the pipeline of all phases from the \as{(input to the)} CFG-to-DAG phase onwards.
\gaurav{Thus, our certificate relates Boogie's VC to the original source AST program \as{so long as}\asout{if} these prior translation \as{steps} are correct.}

\paragraph{CFG decomposition.}

When tackling the certification of \emph{each} phase, we further break down validation of a procedure's CFG in the source program of the phase \gout{in a two-tiered manner:}\gaurav{into sub-problems for each block in the CFG. We prove two results for each block in the source CFG:}

\begin{enumerate}

\item \emph{Local block lemmas:} We prove \gout{independent lemmas per CFG block}\gaurav{an independent lemma for each source CFG block \as{in isolation}}, relating the executions through \gout{a}\gaurav{the} block \gout{in the source program}with the corresponding block\gout{(s)} in the target program \gaurav{(or the VC generated for that block, \as{in the case of}\asout{for} the VC phase)}. In particular, \gout{these}\gaurav{this} lemma\gout{s} \gout{imply}\gaurav{implies} that if the target block\gout{(s)} \gout{have}\gaurav{has} no failing executions (or the VC generated for that block holds, for the VC phase), neither \gout{did}\gaurav{does} the source block \gaurav{for \asout{a related input state}\as{corresponding input states}}.
%

\item \emph{Global block theorems:} We show analogous per-block results concerning all executions \emph{from this block onwards} \gaurav{extending to the end of the \asout{program}\as{procedure in question}}; we build these compositionally by reverse-topological traversal of either the source or target CFGs, as appropriate.
\gaurav{The global block theorem for the entry block establishes correctness of the phase.}
\end{enumerate}

\noindent
This decomposition \gout{over program structure}separates command-level reasoning (local block lemmas) from CFG-level reasoning (global block theorems). It enables concise lemmas and proofs in Isabelle and makes each comprehensible to a human.

\section{A Formal Semantics for Boogie}\label{sec:boogie_lang}
Our certificates prove that the validity of a VC generated by Boogie formally implies correctness of the Boogie \gaurav{CFG-to-DAG source} program\gout{to be verified}. This proof relies crucially on a formal semantics for Boogie itself. Our first contribution is the first such formal semantics for a significant subset of Boogie, mechanised in Isabelle. Our semantics uses the Boogie reference manual~\cite{boogie}, the presentation of its type system~\cite{LeinoR10}, and the Boogie implementation for reference; none of those provide a formal account of the language.
For space reasons, we explain only the key concepts of our detailed formalisation here; \gout{we will make the full Isabelle mechanisation available as part of our accompanying artifact}\gaurav{more details are provided in~\appreftr{A} and the full Isabelle mechanisation is available as part of our accompanying artifact~\cite{artifact}}.

\subsection{The Boogie Language}

Boogie programs consist of a set of top-level declarations of global variables and constants (the \emph{global data}), axioms, uninterpreted (polymorphic) functions, type constructors, and procedures. A procedure declaration includes parameter, local-variable, and result-variable declarations (the \emph{local data}), a pre- and postcondition, and a procedure body given as a CFG.\footnote{Source-level procedure specifications also include \emph{modifies clauses}, declaring a set of global variables the procedure may modify. As we tackle Boogie programs after procedure calls have been desugared, there are no modifies clauses in our formalisation.} CFGs are formalised as usual in terms of basic blocks (containing a possibly-empty list of \emph{basic commands}), and edges; semantically, execution after a basic block continues via any of its successors non-deterministically.

\begin{figure}[t]
    \begin{align*}
    e ::=&\; x \mid \bfalse \mid \btrue \mid i \mid \binaryop{e_1}{\mathit{bop}}{e_2} \mid \unaryop{\mathit{uop}}{e} \mid \funcall{f}{\vec \tau}{\vec e} \mid \old{e} \mid \\
    &\; \bforall{x}{\tau}{e} \mid \bexists{x}{\tau}{e} \mid \bforallt{t}{e} \mid \bexistst{\tau}{e} \\
    \tau ::=&\; \mathit{Int} \mid \mathit{Bool} \mid \tcon{\mathit{C}}{\vec \tau} \mid t \quad
    c ::=\; \assume{e} \mid \assert{e} \mid \assign{x}{e} \mid \havoc{x}
    \end{align*}
    \caption{
        The syntax of our formalised Boogie subset, where $\tau$, $e$, and $c$, denote the types, expressions, and basic commands respectively; control-flow is handled via CFGs over the basic commands. $\mathit{bop}$ and $\mathit{uop}$ denote binary and unary operations, respectively.
        \gout{We assume that procedure calls have been desugared into basic commands.}
    }
    \label{fig:boogie_syntax}
\end{figure}

The types, expressions, and basic commands in our Boogie subset are shown in~\figref{fig:boogie_syntax}.
We support the primitive types $\mathit{Int}$ and $\mathit{Bool}$; types obtained via declared type constructors are \emph{uninterpreted types}; the sets of values such types denote are constrained only via Boogie axioms and \assumeNoArg{} commands.
\gaurav{Moreover, types can contain type variables (for instance, to specify polymorphic functions).}

Boogie expression syntax is largely standard (\eg{} including typical arithmetic and boolean operations).
Old-expressions $\old{e}$ evaluate the expression $e$ w.r.t.\ the current local data and the global data as it \emph{was} in the pre-state of the procedure execution.
Boogie expressions also include universal and existential \emph{value} quantification \gaurav{(\as{written}\asout{\eg{}} $\bforall{x}{\tau}{e}$ \gaurav{and $\bexists{x}{\tau}{e}$})}, as well as universal and existential \emph{type} quantification \gaurav{(\as{written}\asout{\eg{}} $\bforallt{t}{e}$ and $\bexistst{t}{e}$)}. \as{In the latter, $t$ is bound in $e$ and quantifies over \gout{all}\gout{(}\emph{closed}\gout{)} Boogie types \gaurav{(\ie{} types that do not contain any type variables)}.}

Basic commands form the single-steps of traces through a Boogie CFG; sequential composition is implicit in the list\gout{s} of basic commands in a CFG basic block and further control flow (including loops) is prescribed by CFG edges. Boogie's basic commands are assumes, asserts, assignments, and havocs; $\havoc{x}$ non-deterministically assigns a value matching the type of variable $x$ to $x$.

The main Boogie features \emph{not} supported by our subset are maps and other primitive types such as bitvectors.
Boogie maps are polymorphic and impredicative, \ie{} one can define maps that contain themselves in their domain.
Giving a semantic model for such maps in a proof assistant such as Isabelle or Coq is non-trivial; we aim to tackle this issue in the future. Modelling bitvectors will be simpler, although maintaining full automation may require some additional work.

\subsection{Operational Semantics}

\paragraph{Values and state model.}
Our formalisation embeds integer and boolean values shallowly as their Isabelle counterparts; an Isabelle carrier type for all \emph{abstract values} (those of uninterpreted types) is a parameter of our formalisation. Each uninterpreted type is (indirectly) associated with a \emph{non-empty} subset of abstract values via a \gout{surjective}\emph{type interpretation} map $\typs$ from abstract values to (single) types; particular interpretations of uninterpreted types can be obtained via different choices of type interpretation $\typs$.

One can understand Boogie programs in terms of the sets of possible \emph{traces} through each procedure body. Traces are (as usual) composed of sequences of steps according to the semantics of basic commands and paths through the CFG; these can be finite or infinite (representing a non-terminating execution). A trace may halt in three \asout{different }cases: (1)~\gout{the}\gaurav{an} exit block of the procedure is reached in a state satisfying the procedure's postcondition (a \emph{complete} trace),\footnote{The case of the postcondition \emph{not} holding is subsumed under point~(2), since Boogie checks postconditions by generating extra \assertNoArg{} statements.} (2)~an \assert{$A$} command is reached in a state not satisfying assertion $A$ (a \emph{failing} trace), or (3)~an \assume{$A$} command is reached in a state not satisfying $A$ (a trace which \emph{goes to magic} \gaurav{and stops}).
Our formalisation correspondingly includes three kinds of Boogie program states: a distinguished \emph{failure state} \failure{}, a distinguished \emph{magic state} \magic{}, and \emph{normal states}
\normal{($\mathit{os}, \mathit{gs}, \mathit{ls}$)}. A normal state is a triple of partial mappings from variables to values for the old global state (for the evaluation of old-expressions), the (current) global state, and the local state, respectively.

\paragraph{Expression evaluation.}
An expression $e$ evaluates to value $v$ if the \gaurav{(big-step)} judgement $\expreval{\typs}{\Lambda}{\Gamma}{\Omega}{e}{\normal{\mathit{ns}}}{v}$ holds in the context $(\typs,\Lambda,\Gamma,\Omega)$.
Here, $\typs$ is a \emph{type interpretation} (as above), $\Lambda$ is a \emph{variable context}: a pair $(G,L)$ of type declarations for the global ($G$) and local ($L$) data.
$\Gamma$ is a \emph{function interpretation}, which maps each function name to a semantic function mapping a list of types and a list of values to a return value.
The type substitution $\Omega$ maps type variables to types.

The rules defining this judgement can be found in \appreftr{A.2}. For example, the following rule expresses when a universal type quantification evaluates to $\btrue$ \gaurav{($t$ is bound to the quantified type and may occur in $e$)}:
    \begin{align*}
         \Inf{\forall \tau.\; \isclosedty{\tau} \Longrightarrow \expreval{\typs}{\Lambda}{\Gamma}{\Omega(t \mapsto \tau)}{e}{\mathit{ns}}{\btrue{}} }{\expreval{\typs}{\Lambda}{\Gamma}{\Omega}{\bforallt{t}{e}}{\mathit{ns}}{\btrue{}}}
    \end{align*}
\noindent
The premise requires one to show that the expression $e$ reduces to $\btrue$ for every possible type $\tau$ that is closed\gout{(\ie{} does not contain any type variables)}. In general, expression evaluation is  possible only for well-typed expressions; we also formalise Boogie's type system and (for the first time) prove its type safety \gaurav{for expressions in Isabelle}.

\paragraph{Command and CFG reduction.}
The \gaurav{(big-step)} judgement $\redcmd{\typs}{\Lambda}{\Gamma}{\Omega}{c}{s}{s'}$ defines when a command $c$ reduces in state $s$ to state $s'$; the rules are in \appreftr{A.3}. This reduction is lifted to lists of commands $cs$ to model the semantics of a single trace through a CFG block (the judgement $\redcmdlist{\typs}{\Lambda}{\Gamma}{\Omega}{cs}{s}{s'}$).
The operational semantics of CFGs is modelled by the \gaurav{(small-step)} judgement $\redcfg{\typs}{\Lambda}{\Gamma}{\Omega}{G}{\delta}{\delta'}$, expressing that the CFG configuration $\delta$ reduces to configuration $\delta'$ in the CFG $G$.
A CFG configuration is either \emph{active} or \emph{final}.
An active configuration is given by a tuple $(\inl{b_n},s)$, where $b_n$ is the block identifier indicating the current position of the execution and $s$ is the current state.
A final configuration consists of a tuple $(\inr{()},s)$ for state $s$ (and unit value $()$) and is reached at the end of a block that has either no successors, or is in a magic or failure state.

\subsection{Correctness}\label{sec:procedure_correctness}

\begin{figure}[t]
\begin{tikzpicture}[baseline=(current bounding box.north)]
    \tikzset{vertex/.style = {shape=rectangle, draw}}
    \node[align=left](program) at (0,0)
    {
        \assume{\bcode{i != 0}}\\
        \bcode{j := 0}\\
        \bkeyword{while}\bcode{(i != 0)}\\
        \bkeyword{inv}\;\bcode{j >= 0} $\wedge$ \bcode{(i = 0} $\Rightarrow$ \bcode{j > 0)}\\ 
        \bcode{\{} \\
        \quad \bkeyword{if}(\bcode{i < 5}) \bcode{\{}\\
        \quad \quad  \bcode{j := j+1}\\
        \quad \bcode{\}}\\
        \quad \bcode{i := i-1}\\        
        \bcode{\}} \\
        \assert{\bcode{j > 0}}
    };
\end{tikzpicture}
\hfill
\begin{tikzpicture}[baseline=(current bounding box.north)]
    \tikzset{vertex/.style = {shape=rectangle, draw, inner sep=2}}
    \tikzset{edge/.style = {->,> = latex'}}
    
    \node[align=left, vertex, label=right:{$B_0$}](nbefore) at (0,5) {\assume{\bcode{i != 0}}\\
    \bcode{j := 0}};
    \node[align=left, vertex, label=right:{$B_1$}](nhead) at (0,4) {
        \assert{\bcode{j >= 0} $\wedge$ \bcode{(i = 0} $\Rightarrow$ \bcode{j > 0)}}
    };
    \node[align=left, vertex, label=right:{$B_2$}](nbody) at (-1.5,3) {
        \bcodestandard{\assume{\bcode{i != 0}}}
    };
    \node[align=left, vertex, label=right:{$B_3$}](nleft) at (-1.5,2) {
        \assume{\bcode{i < 5}} \\
        \bcode{j := j+1}
    };
    \node[align=left, vertex, label=right:{$B_4$}](nright) at (1.6,1.8) {
        \assume{\bcode{!(i < 5)}}
    };
    \node[align=left, vertex, label=right:{$B_5$}](njoin) at (-1.5,1) {
        \bcode{i := i-1}
    };
    \node[align=left, vertex, label=right:{$B_6$}](nafter) at (1.7,3) {
        \assume{\bcode{i = 0}} \\
        \assert{\bcode{j > 0}}
    };

    \path
    (nbefore) [->] edge node [above] {} (nhead)
    (nhead) [->] edge [above] node [above] {} (nbody)
    (nhead) [->] edge node [above] {} (nafter)
    (nbody) [->] edge node [above] {} (nleft)
    (nbody) [->] edge node [above] {} (nright)
    (nleft) [->] edge node [above] {} (njoin)
    (nright) [->] edge node [above] {} (njoin)
    (njoin.west) [->] edge [bend left=30] node [above] {} (nhead.west);
\end{tikzpicture}
\caption{Running example in source code and CFG representation, respectively.}
\label{fig:running_example_ast_cfg}
\end{figure}
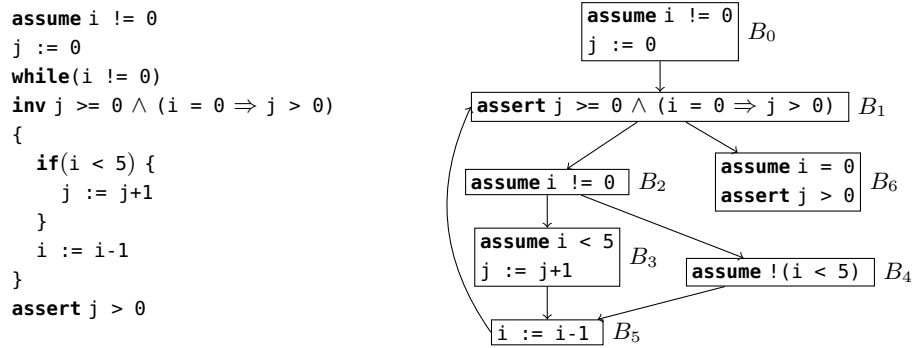

A procedure is \emph{correct} if it has \emph{no failing traces}. This is a \emph{partial correctness} semantics; a procedure body whose traces never leave a loop is trivially correct provided that no intermediate \assertNoArg{} commands fail. Procedure correctness relies on CFG correctness. A CFG $G$ is correct w.r.t.\ a postcondition $Q$ and a context $(\typs,\Lambda,\Gamma,\Omega)$ in an initial normal state $\normal{\mathit{ns}}$ if the following holds for all configurations $(r,s')$:
\begin{align*}
&\redcfgmulti{\typs}{\Lambda}{\Gamma}{\Omega}{G}{(\inl{\entry{G}}, \normal{ns})}{(r,s')} \Longrightarrow [ s' \neq \failure{} \wedge {} \\
&\quad (r = \inr{()} \Longrightarrow (\forall \mathit{ns'}.\; s' = \normal{\mathit{ns'}} \Longrightarrow
\expreval{\typs}{\Lambda}{\Gamma}{\Omega}{Q}{\normal{\mathit{ns'}}}{\btrue{}}
)) ]
\end{align*}
\noindent
where \entry{$G$} is the entry block of $G$ and $\redcfgmultiArrow$ is the reflexive-transitive closure of the CFG reduction. The postcondition is needed only if a final configuration is reached in a normal state, while failing states must be unreachable.
Whenever we omit $Q$, we implicitly mean the postcondition to be simply $\btrue{}$.
In our tool, we consider only empty initial mappings $\Omega$, since we do not support procedure type parameters (lifting our work to this feature will be straightforward).

For a procedure $p$ to be correct w.r.t.\ a \gout{particular}context, its body CFG must be correct w.r.t.\ the same context and $p$'s postcondition, \emph{for all} initial normal states $\normal{ns}$ that satisfy $p$'s precondition and which respect the context.
For $\mathit{ns}$ to \emph{respect} a context, it must be well-typed and must satisfy the axioms when restricted to its constants.
We say that $p$ is \emph{correct}, if it is correct \wrt{} \emph{all well-formed contexts}, which\gout{, among other things,} must have a well-typed function interpretation \gaurav{and a type interpretation that inhabits every uninterpreted closed type (and only those).}

\paragraph{Running example.}

We will use the simple CFG of \figref{fig:running_example_ast_cfg} as a running example, intended as body of a procedure with trivial (\btrue{}) pre- and post-conditions. The code includes a simple loop with a declared loop invariant, which functions as a classical Floyd/Hoare-style inductive invariant, and for the moment can be considered as an implicit \assertNoArg{} statement at the loop head. The CFG has infinite traces: those which start from any state in which \bcode{i} is negative. Traces starting from a state in which \bcode{i} is zero go to magic; they do not reach the loop. The program is correct (has no failing traces): all other initial states will result in traces that satisfy the loop invariant and the \gaurav{final} \assertNoArg{} statement. If we removed the initial \assumeNoArg{} statement, however, there \emph{would} be failing traces: the loop invariant check would fail if \bcode{i} were initially zero.

\section{The CFG-to-DAG Phase}\label{sec:cfg_to_dag}
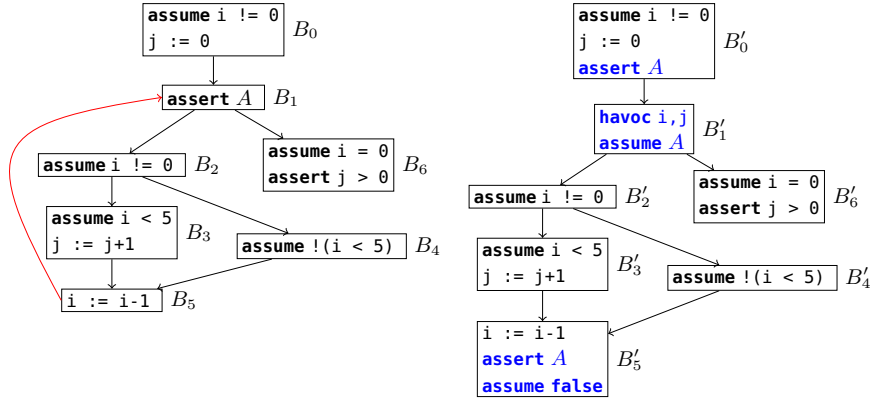
\begin{figure}[t]
    \scalebox{0.9}{
    \begin{tikzpicture}[baseline=(current bounding box.north)]
        \tikzset{vertex/.style = {shape=rectangle, draw, inner sep=2}}
        \tikzset{edge/.style = {->,> = latex'}}
        
        \node[align=left, vertex, label=right:{$B_0$}](nbefore) at (0,5) {\assume{\bcode{i != 0}}\\
        \bcode{j := 0}};
        \node[align=left, vertex, label=right:{$B_1$}](nhead) at (0,4) {
            \assert{$A$}
        };
        \node[align=left, vertex, label=right:{$B_2$}](nbody) at (-1.5,3) {
            \bcodestandard{\assume{\bcode{i != 0}}}
        };
        \node[align=left, vertex, label=right:{$B_3$}](nleft) at (-1.5,2) {
            \assume{\bcode{i < 5}} \\
            \bcode{j := j+1}
        };
        \node[align=left, vertex, label=right:{$B_4$}](nright) at (1.6,1.8) {
            \assume{\bcode{!(i < 5)}}
        };
        \node[align=left, vertex, label=right:{$B_5$}](njoin) at (-1.5,1) {
            \bcode{i := i-1}
        };
        \node[align=left, vertex, label=right:{$B_6$}](nafter) at (1.7,3) {
            \assume{\bcode{i = 0}} \\
            \assert{\bcode{j > 0}}
        };
    
        \path
        (nbefore) [->] edge node [above] {} (nhead)
        (nhead) [->] edge [above] node [above] {} (nbody)
        (nhead) [->] edge node [above] {} (nafter)
        (nbody) [->] edge node [above] {} (nleft)
        (nbody) [->] edge node [above] {} (nright)
        (nleft) [->] edge node [above] {} (njoin)
        (nright) [->] edge node [above] {} (njoin)
        (njoin.west) [->] edge [color=red, bend left=55, looseness=2] node [above] {} (nhead.west);
    \end{tikzpicture}
    }
    \hfill
    \scalebox{0.9}{
    \begin{tikzpicture}[baseline=(current bounding box.north)]
        \tikzset{vertex/.style = {shape=rectangle, draw, inner sep=2}}
        \tikzset{edge/.style = {->,> = latex'}}
        
        \node[align=left, vertex, label=right:{$B'_0$}](nbefore) at (0,5.3) {\assume{\bcode{i != 0}}\\
        \bcode{j := 0} \\
        {\color{NewCmdColor}\assert{$A$}}};
        \node[align=left, vertex, label=right:{$B'_1$}](nhead) at (0,4) {
            {\color{NewCmdColor}\havoc{\bcode{i,j}}} \\
            {\color{NewCmdColor}\assume{$A$}}
        };
        \node[align=left, vertex, label=right:{$B'_2$}](nbody) at (-1.5,3) {
            \bcodestandard{\assume{\bcode{i != 0}}}
        };
        \node[align=left, vertex, label=right:{$B'_3$}](nleft) at (-1.5,2) {
            \assume{\bcode{i < 5}} \\
            \bcode{j := j+1}
        };
        \node[align=left, vertex, label=right:{$B'_4$}](nright) at (1.6,1.8) {
            \assume{\bcode{!(i < 5)}}
        };
        \node[align=left, vertex, label=right:{$B'_5$}](njoin) at (-1.5,0.6) {
            \bcode{i := i-1} \\
            {\color{NewCmdColor}\assert{$A$}} \\
            {\color{NewCmdColor}\assume{\bfalse{}}}
        };
        \node[align=left, vertex, label=right:{$B'_6$}](nafter) at (1.7,3) {
            \assume{\bcode{i = 0}} \\
            \assert{\bcode{j > 0}}
        };
    
        \path
        (nbefore) [->] edge node [above] {} (nhead)
        (nhead) [->] edge [above] node [above] {} (nbody)
        (nhead) [->] edge node [above] {} (nafter)
        (nbody) [->] edge node [above] {} (nleft)
        (nbody) [->] edge node [above] {} (nright)
        (nleft) [->] edge node [above] {} (njoin)
        (nright) [->] edge node [above] {} (njoin);
    \end{tikzpicture}
    }
    \caption{The CFG-to-DAG phase applied to the running example (source is left, target is right). The back-edge (the red edge \gaurav{from $B_5$ to $B_1$} in the left CFG) is eliminated. The blue commands are new. $A$ is given by \bcode{j >= 0} $\wedge$ \bcode{(i = 0} $\Rightarrow$ \bcode{j > 0)}.}
    \label{fig:cfg_to_dag_example}
\end{figure}

In this section, we present the validation for the CFG-to-DAG phase in the Boogie verifier. This phase is challenging as it changes the CFG structure, inserts additional non-deterministic assignments and \assumeNoArg{} statements, and must do so correctly for arbitrary (reducible) nested loop structures\gaurav{, which can include unstructured control flow (\eg{} jumps out of loops)}.

\subsection{CFG-to-DAG Phase Overview}

The CFG-to-DAG phase applies to every \emph{loop head} block identified by Boogie's implementation and any \emph{back-edges} from a block reachable from the loop head block back to the loop head (following standard definitions for reducible CFGs~\cite{HechtU72}).
\figref{fig:cfg_to_dag_example} illustrates the phase's effect  on our \asout{simple }running example. Block $B_1$ is the (only) loop head here, and the edge from $B_5$ to it \asout{is }the only back-edge \asout{in this program }(completing looping paths via $B_2$ and $B_3$ or $B_2$ and $B_4$). An \assert{$A$} statement starting a loop head (like $B_1$) is interpreted as declaring $A$ to be the loop invariant.\footnote{In general, multiple \as{asserts}\asout{assertions} at the beginning of a loop head \as{may} form the invariant.\asout{ We focus on a single assertion here for simplicity.}} The CFG-to-DAG phase performs the following steps:

\begin{enumerate}
\item\label{stepi} Accumulate a set $\modX$ of all (local and global) variables \emph{assigned-to} on \emph{any looping path} from the loop head back to itself. In our example, $\modX$ is $\{i,j\}$.
\item\label{stepii} Move the \assert{$A$} statement declaring a loop invariant (if any) from the loop head to the end of \emph{each preceding} block (in our example: $B_0$ and $B_5$).
\item\label{stepiii} Insert \havocNoArg{} statements at the start of the loop head block per variable in $\modX$, followed by a single \assume{$A$} statement (preceding any further statements).
\item\label{stepiv} For each block with a back-edge to the loop head, delete the back-edge; if this leaves the block with no successors, append \assume{\bfalse{}} to its commands.%
\footnote{\gaurav{Omitting \assume{\bfalse{}} if there are no successors would be incomplete, since otherwise the postcondition would have to be satisfied.}}
\end{enumerate}

The havoc-then-assume sequence introduced in step \ref{stepiii} can be understood as generating traces for \emph{arbitrary values of $\modX$} satisfying the loop invariant $A$, effectively over-approximating the set of states reachable at the loop head in the original program. In particular, the remnants of any originally looping path (\eg{} $B'_1$,$B_2'$,$B_3'$,$B'_5$) enforce that any non-failing trace starting from any such state must (due to the \assertNoArg{} added to block $B'_5$ in step \ref{stepii}) result in a state which re-establishes the loop invariant. Such paths exist only to enforce this inductive step (analogously to the premise of a Hoare logic while rule); so long as the \assertNoArg{} succeeds, we can discard these traces via step \ref{stepiv}.

While we illustrate this step on a simple CFG, in general a loop head may have multiple back-edges, looping structures may nest, and edges may exit multiple loops. For the above translation to be correct, the CFG must be reducible and loop heads and corresponding back-edges identified accurately, which is complex in general. Importantly (but perhaps surprisingly), our work makes this phase of Boogie certifying \emph{without} explicitly verifying (or even defining) these notions.

\subsection{CFG-to-DAG Certification: Local Block Lemmas}

We define first our local block lemmas for this phase. Recall that these prove that if executing the statements of a target block yields no failing executions, the same holds for the corresponding source block; this result is trivial for source blocks other than loop heads and their immediate predecessors, since these are unchanged in this phase. To enable eventual composition of our block lemmas, we need to also reflect the role of the \assumeNoArg{} and \assertNoArg{} statements employed in this phase. The formal statement of our local block lemmas is as follows\footnote{We omit some details regarding well-typedness, handled fully in our formalisation.}:

\begin{theorem}[CFG-to-DAG Local Block Lemma]
\gaurav{
Let $B$ be a source block with commands $\srcBlock$, whose corresponding target block has commands $\trgBlock$.}
\gout{Let $\srcBlock$ and $\trgBlock$ be corresponding source and target blocks, respectively, for the CFG-to-DAG transformation.}If \gout{$\srcBlock$}\gaurav{$B$} is a loop head, let $\modX$ be as defined in CFG-to-DAG step \ref{stepi} (and empty otherwise) and let $A_\mathit{pre}$ be its loop invariant (or $\btrue{}$ otherwise). If \gout{$\srcBlock$}\gaurav{$B$} is a \emph{predecessor} of a loop head, let $A_\mathit{post}$ be the loop invariant of its successor (and $\btrue{}$ otherwise). Then, \emph{if}:
\begin{enumerate}
\item \redcmdlist{\typs}{\Lambda}{\Gamma}{\Omega}{\srcBlock}{\normal{\mathit{ns}_1}}{s_1'}
\item $\forall s_2'.\; \redcmdlist{\typs}{\Lambda}{\Gamma}{\Omega}{\trgBlock}{\normal{\mathit{ns}_2}}{s_2'} \Longrightarrow s_2' \neq \failure{}$
\item $A_\mathit{pre}$ is satisfied in $\mathit{ns}_1$, and $\mathit{ns}_2$ differs from $\mathit{ns}_1$ only on variables in $X_H$ and variables not defined in $\Lambda$
\end{enumerate}
\emph{then}:
$s_1' \neq \failure{}$ and if $s_1'$ is a normal state, then (1) $A_\mathit{post}$ is satisfied in $s_1'$, and (2) \gaurav{if no \assume{\bfalse} was added at the end of $\trgBlock$, then} there is a target execution in $cs_T$ from $\normal{ns_2}$ that reaches a normal state that \gout{does not differ from $s_1'$ on any variables other than those \gout{not}defined in $\Lambda$}\gaurav{differs from $s_1'$ only on variables not defined in $\Lambda$}.
\label{thm:cfgtodag_local}
\end{theorem}

The gist of this lemma is to capture \emph{locally} the ideas behind the four steps of the phase. For example, consequence~(1) reflects that \emph{after} the transformation, any blocks that \emph{were previously} predecessors of a loop head ($B'_0$ and $B'_5$ in our running example) will have an \assertNoArg{} statement checking for the corresponding invariant (and so if the target program has no failing traces, in each trace this invariant will be true at that point).

\subsection{CFG-to-DAG Certification: Global Block Theorems}

We lift our certification to \emph{all} traces through the source and target CFGs; the statement of the corresponding global block theorems is similar to that of local block theorems lifted to CFG executions, and for space reasons we do not present it here, but it is included in our Isabelle formalisation. In particular, we prove for each block (working in reverse topological order through the target CFG blocks) that if executions starting in the target CFG block never fail, neither do any executions starting from the corresponding source CFG block, and looping paths modify at most the variables havoced according to step \ref{stepiii} of the phase.

The major challenge in these proofs is reasoning about looping paths in the source CFG, since these revisit blocks. To solve this challenge, we perform inductive arguments per loop head in terms of the number of steps remaining in the trace in question.\footnote{This may seem insufficient since traces can be infinite, but importantly a \emph{failing} trace is always finite, and our theorems need only eliminate the chance of failing traces.}
Our global block theorem for a block $B$ then carries as an assumption an induction hypothesis for each loop that contains $B$. Proving a global block theorem for the origin of a back-edge is taken care of by applying the corresponding induction hypothesis.

This proof strategy works only if we have obtained the induction hypothesis for the loop head \emph{before} we use the global block theorem of the origin of a back-edge (otherwise we cannot discharge the block theorem's hypothesis).
In other words, our proof implicitly shows the necessary requirement that loop heads (as identified by Boogie) dominate all back-edges reaching them \emph{without us formalising any notion of domination, CFG reducibility, or any other advanced graph-theoretic concept}. This shows a major benefit of our validation approach over a once-and-for-all verification of Boogie itself: our proofs indirectly check that the identification of loop heads and back-edges guarantees the necessary \emph{semantic properties} without being concerned with \emph{how} Boogie's implementation computes this information.

Our approach applies equally to nested loops and more-generally to reducible CFG structures; \emph{all} corresponding induction hypotheses are carried through from the visited loop heads. The requirement that no more than the havoced variables $X_H$ are modified in the source program is easily handled by showing that variables modified in an inner loop are a subset of those in outer loops. As for all of our results, our global block lemmas are proven automatically in Isabelle per Boogie procedure, providing per-run certificates for this phase.

\section{The Passification Phase}\label{sec:passification}
\begin{figure}[t]
\scalebox{0.92}{
    \begin{tikzpicture}[baseline=(current bounding box.north)]
    \tikzset{vertex/.style = {shape=rectangle, draw, inner sep=2}}
    \tikzset{edge/.style = {->,> = latex'}}
    
    \node[align=left, vertex, label=right:{$B'_2$}](nbody) at (0,2.3) {
        \bcodestandard{\assume{\bcode{i != 0}}}
    };
    \node[align=left, vertex, label=right:{$B'_3$}](nleft) at (-1.5,1.3) {
        \assume{\bcode{i < 5}} \\
        \bcode{j := j+1}
    };
    \node[align=left, vertex, label=right:{$B'_4$}](nright) at (1.6,1.4) {
        \assume{\bcode{!(i < 5)}}
    };
    \node[align=left, vertex, label=right:{$B'_5$}](njoin) at (0.1,0) {
        \bcode{i := i-1} \\
        {\assert{\bcode{j >= 0} $\wedge$ \bcode{(i = 0} $\Rightarrow$ \bcode{j > 0)}}} \\
        {\assume{\bfalse{}}}
    };

    \path
    (nbody) [->] edge node [above] {} (nleft)
    (nbody) [->] edge node [above] {} (nright)
    (nleft) [->] edge node [above] {} (njoin)
    (nright) [->] edge node [above] {} (njoin);
\end{tikzpicture}
}
\scalebox{0.92}{
    \begin{tikzpicture}[baseline=(current bounding box.north)]
        \tikzset{vertex/.style = {shape=rectangle, draw, inner sep=2}}
        \tikzset{edge/.style = {->,> = latex'}}
        
        \node[align=left, vertex, label=right:{$B''_2$}](nbody) at (0,2.5) {
            \bcodestandard{\assume{\bcode{i1 != 0}}}
        };
        \node[align=left, vertex, label=right:{$B''_3$}](nleft) at (-1.5,1.5) {
            \assume{\bcode{i1 < 5}} \\
            \assume{\bcode{j3 = j2+1}} \\
            {\color{darkgreen}\assume{\bcode{j4 = j3}}}
        };
        \node[align=left, vertex, label=right:{$B''_4$}](nright) at (2.1,1.4) {
            \assume{\bcode{!(i1 < 5)}} \\
            {\color{darkgreen}\assume{\bcode{j4 = j2}}}
        };
        \node[align=left, vertex, label=right:{$B''_5$}](njoin) at (0.1,0) {
            \assume{\bcode{i2 = i1-1}} \\
            {\assert{\bcode{j4 >= 0} $\wedge$ \bcode{(i2 = 0} $\Rightarrow$ \bcode{j4 > 0)}}} \\
            {\assume{\bfalse{}}}
        };
    
        \path
        (nbody) [->] edge node [above] {} (nleft)
        (nbody) [->] edge node [above] {} (nright)
        (nleft) [->] edge node [above] {} (njoin)
        (nright) [->] edge node [above] {} (njoin);
    \end{tikzpicture}
}
\caption{The passification phase applied to the branch in the running example with the result on the right. The \gaurav{final (green) commands in $B_3''$ and $B_4''$} \gout{green commands} are the synchronisation commands. At the uppermost blocks shown here, the current versions of \bcode{i} and \bcode{j} are \bcode{i1} and \bcode{j2}, respectively. The full CFGs are shown\gout{in \figref{fig:running_example_dag_full_app} and \figref{fig:running_example_passive_full_app}} in \appreftr{B}.}
\label{fig:passification_example}
\end{figure}
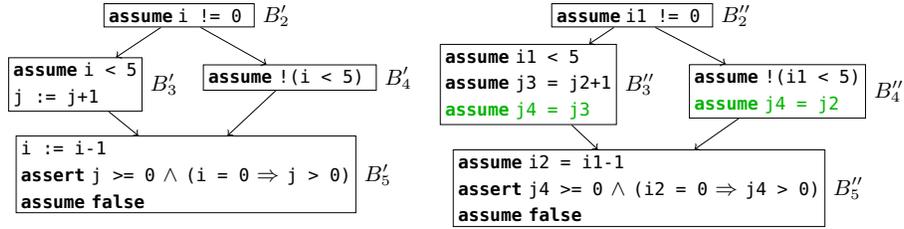

In this section, we describe the validation of the passification phase in the Boogie verifier. Unlike the previous phase, passification makes no changes to the CFG structure, but makes substantial changes to the program states (via SSA-like renamings), substantially increases non-determinism, and employs \assumeNoArg{} statements to re-tame the sets of possible traces.

\subsection{Passification Phase Overview}
\label{sec:passification_details}

The main goal of passification is to eliminate assignments such that a more efficient VC can be ultimately generated \cite{FlanaganSaxe2001,LeinoEfficientWp05,BarnettLeino2005}. In the Boogie verifier, this is implemented as a single transformation phase that can be thought of as two independent steps. Firstly, the source CFG is transformed into \emph{static single assignment} (SSA) form, introducing \emph{versions} (fresh variables) for each original program variable such that each version is assigned at most once in any program trace. In a second step, variable assignments are \emph{completely eliminated}: each assignment command $\assign{x}{e}$ is replaced by $\assume{x=e}$. Havoc statements are simply removed; their effect is implicit in the fact that a new variable version is used (via the SSA step) \emph{after} such a statement.

\figref{fig:passification_example} shows the effect of this phase on four blocks of our running example (the full figure of the target CFG is shown \gout{in \figref{fig:running_example_passive_full_app}}in \appreftr{B}). The commands \gout{highlighted in green are}inserted just before \gaurav{the} join \gaurav{block (here, $B_5''$)} \gout{in the CFG structure to}introduce a consistent variable version (here, \bcode{j4}) for use in the \gout{subsequent}\gaurav{join} block.
It is convenient to speak of target variables in terms of their source program counterparts: we say \eg{} that \bcode{j} \emph{has version $4$} on entry to block $B'_5$.

Compared to traces through the source program, the space of variable values in a trace through the target program is initially much larger; each version may, on entry to the CFG, have an arbitrary value. For example, \bcode{j4} may have any value on entry to $B_2''$; traces in which its value does not correspond to the constraint of the \assumeNoArg{} statements in $B_3''$ or $B_4''$ will go to magic and not reach $B_5''$. Importantly, however, not \emph{all} traces go to magic; enough are preserved to simulate the executions of the original program: each \assumeNoArg{} statement constrains the value of exactly one variable version, and the same version is never constrained more than once.
Capturing this delicate argument formally is the main challenge in certifying this step.

As extra parts of the passification phase, the Boogie verifier performs constant propagation and  desugars old-expressions (using variable versions appropriate to the entry point of the CFG).
We omit their descriptions here for brevity, but our implementation certifies them.

\subsection{Passification Certification: Local Block Lemmas}
To validate the passification phase, it is sufficient to show that each source execution is simulated by a corresponding target execution, made precise by constructing a relation between the states in these executions. Such \emph{forward simulation} arguments are standard for proving correctness of compilers for deterministic languages. However, the situation here is more complex due to the fact that the target CFG has a much wider space of traces: the values of each versioned variable in the target program are initially unconstrained, meaning traces exist for all of their combinations. On the other hand, many of these traces do not survive the \assumeNoArg{} statements encountered in the target program. Picking the correct \emph{single} trace or state to simulate a particular source execution would require knowledge of all variable assignments that are \emph{going} to happen, which is not possible due to non-determinism and would preclude the block-modular proof strategies that our validation approach employs.

Instead, we generalise this idea to relating each single source state $s$ with a \emph{set} $T$ of corresponding target program states. We define variable relations $\varRel$ at each point in a trace, making explicit the mappings used in the SSA step between source program variables and their corresponding versions. For example, on entry to block $B'_2$ in the source version of our running example (correspondingly $B''_2$ in the target), the $\varRel$ relation relates \bcode{i} to \bcode{i1} and \bcode{j} to \bcode{j2}. All states $t\in T$ must precisely agree with $s$ w.r.t.\ $\varRel$ (\eg{}, $s(\bcode{i}) = t(\bcode{i1})$, $s(\bcode{j}) = t(\bcode{j2})$). On the other hand, our sets of states $T$ are defined to be completely unconstrained (besides typing) for \emph{future} variable versions. For example, for every $t\in T$ at the same point in our example, there will be states in $T$ assigning each possible value (of the same type) to \bcode{i2} (and otherwise agreeing with $t$).

More precisely, for a set of variables $X$, we say that a set of states $T$ \emph{constrains at most} $X$ \gaurav{\emph{w.r.t. variable context $\Lambda$}} if, for every $t \in T$, $z \notin X$, \gaurav{$z$ is in $\Lambda$,} and value $v$ of $z$'s type, we have $t[z\mapsto v]\in T$. In other words, the set $T$ is closed under arbitrary changes to values of all variables in \gaurav{$\Lambda$ but }\emph{not} in $X$. We construct our sets $T$ such that they constrain at most \emph{current and past versions} of program variables. It is this fact that enables us to handle subsequent \assumeNoArg{} statements in the target program and, in particular, to show that the set of possible traces in the target program never becomes empty while there are possible traces in the source program. For example, when relating the source command \bcode{j := j+1} in $B'_3$ with the target command \assume{\bcode{j3 = j2 + 1}} in block $B''_3$, we use the fact that our set of states does not constrain \bcode{j3} to prove that, although many traces go to magic at this point, for a non-empty set of states $T'\subseteq T$ (those in which \bcode{j3} has the ``right'' value equal to \bcode{j2 + 1}), execution continues in the target.

We now make these notions more precise by showing the definition of our local block lemmas for the passification phase\footnote{We omit some details regarding well-typedness, handled fully in our formalisation.}.

\begin{theorem}[Passification Local Block Lemma]
Let $B$ be a source block with commands \srcCmds, whose corresponding target block has commands \trgCmds; let $\varRel$ and $\varRel'$ be the variable relations at the beginning and end of $B$, respectively.  Let $X$ be a set of variable versions, and $\normal{\mathit{ns}}$ be a normal state. Let $T$ be a non-empty set of normal states such that $\normal{\mathit{ns}}$ agrees with $T$ according to $\varRel$, and $T$ constrains at most $X$ \gaurav{w.r.t. $\Lambda_2$}.
Furthermore, let $Y$ be the variable versions corresponding to the targets of assignment and havoc statements in \srcCmds. If both
\begin{enumerate}
\item $\redcmdlist{A}{\Lambda_1}{\Gamma}{\Omega}{\srcCmds}{\normal{\mathit{ns}}}{s'} \wedge \gaurav{s' \neq \magic{}}$
\item\label{condii} $X \cap Y = \emptyset$
\end{enumerate}
then there exists a non-empty set of normal states $T' \subseteq T$ s.t. $T'$ \gout{depends only on}\gaurav{constrains at most} $X \uplus Y$ \gaurav{w.r.t. $\Lambda_2$} and for each \gout{normal state}$t' \in T'$, there exists a state $t'^*$ s.t.
\begin{enumerate}
    \item $\redcmdlist{A}{\Lambda_2}{\Gamma}{\Omega}{\mathit{cs}_2}{t'}{t'^*} \wedge (s' = \failure{} \Longrightarrow t'^* = \failure{})$
    \item If $s'$ is a normal state, then $s'$ and $t'$ are related w.r.t. $\mathcal{V}_R'$ (and $t'^* = t'$).
\end{enumerate}
\label{thm:passification_local}
\end{theorem}
This lemma captures our generalised notion of forward simulation appropriately. The first conclusion expresses that the target does not get stuck and that failures are preserved, while the second shows that if \gaurav{the source} execution neither fails nor stops then the resulting states are related. Note that premise \ref{condii} is essential in the proof to guarantee that the \assumeNoArg{} statements introduced by passification do not eliminate the chance to simulate source executions; the condition expresses that the variable versions newly constrained do not intersect with those previously constrained. To prove these lemmas over the commands in a single block, we are forced to check that the same version is not constrained twice.

\subsection{Passification Certification: Global Block Theorems}
As for all phases, we lift our local block lemmas to theorems certifying all executions \emph{starting} from a particular block, and thus, ultimately, to entire CFGs. For the passification phase, most of the conceptual challenges are analogous to those of the local block lemmas; we similarly employ $\varRel$ relations between source variables and their corresponding target versions. To connect with our local block lemmas (and build up our global block theorems, which we do backwards through the CFG structure), we repeatedly require the key property that the set of variable versions constrained in our executions so far is disjoint from those which may be constrained by a subsequent \assumeNoArg{} statement (\cf{} premise \ref{condii} of our local block lemma above). Concretely tracking and checking disjointness of these concrete sets of variables is simple, but turns out to get expensive in Isabelle when the sets are large.

We circumvent this issue with our own \emph{global versioning scheme} (as opposed to the versions used by Boogie, which are \emph{independent} for different source variables): according to the CFG structure, we assign a \emph{global} version number
\globversion{x} to each variable $x$ in the target program such that, if $x$ is constrained in a target block $B'$ and $y$ is constrained in another target block $B''$ reachable from $B'$, then $\globversion{x} < \globversion{y}$. Such a consistent global versioning always exists in the target programs generated by Boogie because the only variables not constrained exactly once \emph{in the program} are those used to synchronise executions (\ie{} \bcode{j4} in~\figref{fig:passification_example}), which always appear right before branches are merged.
We can now encode our disjointness properties \gout{(which imply this fact)}much more cheaply: we simply compare the \emph{maximal} global version of all already-constrained variables with the \emph{minimal} global version of those (potentially) to be constrained.
Since we represent variables as integers in the mechanisation, we directly use our global version \emph{as} the variable name for the target program; there is no need for an extra lookup table.
Note that (readability aside) it makes no difference which variables names are used in intermediate CFGs; 
we ultimately care only about validating the original CFG\@.

\section{The VC Phase}\label{sec:vc_phase}
In this section, we present the validation of the VC phase in the Boogie verifier. This phase has two main aspects: (1)~it encodes and desugars all aspects of the Boogie type system, employing additional uninterpreted functions and axioms to express its properties~\cite{LeinoR10}; program expression elements such as Boogie functions are analogously desugared in terms of these additional uninterpreted functions, creating a non-trivial logical gap between expressions as represented in the VC and those from the input program. (2)~It performs an efficient (block-by-block) calculation of a weakest precondition for the (acyclic, passified) CFG, resulting in a formula characterising its verification requirements, subject to background axioms and other hypotheses.

\subsection{VC Structure}\label{sec:vc_components}
The generated VC has the following overall structure (represented as a shallow embedding in our certificates)\footnote{Note that top-level quantification over functions is implicit in the (first-order) SMT problem generated by Boogie; we quantify explicitly in our Isabelle representation.}:
\begin{align*}
\forall \underbrace{\textit{VC quantifiers}}_{
    \begin{subarray}{l}
     \text{type encoding parameters,} \\
     \text{functions, variable values}
    \end{subarray}
    }.
    \quad
     (\underbrace{\textit{VC assumptions}}_{
     \begin{subarray}{l}
        \text{type encoding,} \\
       \text{func./var./prog. axioms}
     \end{subarray}
     } \Longrightarrow \textit{CFG WP})
\end{align*}
The VC quantifies over parameters required for the type encoding, as well as VC counterparts representing the variable values and functions in the Boogie program.
The VC \gaurav{body} is an implication\gout{:}\gaurav{, \as{whose}\asout{where the}} premise contains: \gaurav{(1)} assumptions that axiomatise the \gout{functions generated in this phase to desugar the type system}\gaurav{type encoding parameters},  \gaurav{(2)} axioms expressing the typing of \gaurav{Boogie} variables and functions, \gout{as well as}\gaurav{and (3)} assumptions directly relating to axioms explicitly declared in the Boogie program. The conclusion of the implication is an optimised version of the weakest (liberal) precondition (WP) of the CFG.\footnote{One difference in our version of the Boogie verifier is that we switched off the generation of extra variables introduced to report error traces~\cite{LeinoMillsteinSaxe2005}; these are redundant for programs that do not fail and further complicate the VC structure.}
%


\subsection{Boogie's Logical Encoding of the Boogie Type System}\label{sec:typesystem_desugaring}
We first briefly explain Boogie's logical encoding of its own type system. Values and types are represented at the VC level by two uninterpreted carrier sorts $V$ and $T$. An uninterpreted function $\typeofvalNoArg$ from $V$ to $T$ maps each value to the representation of its type. Boogie type constructors are each modelled with an (injective) uninterpreted function $C$ with return sort $T$ and taking arguments (per constructor parameter) of sort $T$. For example, a type constructor \tcon{\textit{List}}{$t$} is represented by a VC function from $T$ to $T$. \gout{Inverse}Projection functions are also generated \gaurav{for each type constructor} ($C^{\pi}_i$ for each type argument at position $i$), \eg{} mapping the representation of a type \tcon{\textit{List}}{$t$} to the representation of type $t$.

\gout{This encoding is then used throughout the Boogie program to map all typed Boogie expressions to untyped VC expressions with types as explicit values.}%
\gaurav{This encoding is then used in the VC to recover Boogie typing constraints for the untyped VC terms.
Recovering the constraints is not always straightforward due to optimisations performed by Boogie.
For example, the VC translation of the Boogie expression $\bforallt{t}{\bforall{x}{List(t)}{e}}$ \asout{($\bforallt{t}{e'}$ quantifies over all types where $t$ is bound to the type)}\as{no longer quantifies}\asout{does not quantify} over types\as{; all original}\asout{ and replaces} occurrences of $t$ \as{in $e$ having been \gout{replaced}\gaurav{translated to} }\asout{in the VC translation of $e$ }
\gout{by}$\textit{List}^{\pi}_1(\typeofval{}{x})$.
This optimisation reflects that this particular type quantification is redundant, since $t$ can be recovered from the type of $x$.\footnote{\gaurav{Note that in the VC the quantification over $x$ ranges over all values of sort $V$. An implication is used to consider only those $x$ for which $\typeofval{}{x} = \textit{List}(\textit{List}^{\pi}_1(\typeofval{}{x}))$.}}}

\gout{This can have a non-trivial effect on the corresponding program elements. For example, a polymorphic Boogie function declared as: \bcode{function foo<t>(x:List t): t} would, in our semantics for Boogie, be a partial function $f$ of type $\mathit{ty} \rightarrow \val{} \rightharpoonup \val{}$, where $f(\tau, v)$ is defined only if $v$ has type \tcon{\textit{List}}{$\tau$} and \gaurav{$\mathit{ty}$ contains no type variables}.
By contrast, the corresponding VC-level function $h_\mathit{vc}$ is \emph{total} of type $\val{} \rightarrow \val{}$; it does not take a type as input (even though this type defines the return type). This modelling suffices because \emph{after this desugaring}, the type parameter is technically redundant: one can recover the return type from the argument value: $\textit{List}^{\pi}_1(\typeofval{}{v})$.}

\subsection{Working from VC Validity}

Our certificates assume that the generated VC is valid (\asout{recall that }certifying the validity-checking of the VC by an SMT solver is an orthogonal concern). However, connecting VC validity back to block\asout{ (and command)}-level properties about the specific program requires a number of technical steps.
\gaurav{We need to construct Isabelle-level semantic values to \emph{instantiate} the top-level quantifiers in the VC such that the corresponding VC assumptions (left-hand side of the VC) can be proved and, thus, validity of the corresponding WP can be deduced. Moreover, we must ensure that our instantiation yields a WP whose validity implies correctness of the Boogie program. For example, a top-level VC quantifier modelling a Boogie function $f$ \as{must}\asout{should} be instantiated \as{with a mathematical function\gout{satisfying $f$'s definition.}} that behaves in the same way as $f$ for arguments of the correct type.}

\gout{Firstly, we need to construct Isabelle-level semantic values (\eg{} functions) to \emph{instantiate} the top-level quantifiers (\eg{} over functions) in the VC\@.}%
We instantiate the carrier sort $V$ for values in the VC with the corresponding type \gout{$\val{}$}denoting Boogie values in our formalisation; the carrier sort $T$ for \emph{types} is instantiated to be all \gout{\emph{closed}}Boogie types \gout{$\closedty{}$}\gaurav{that do not contain free variables (\ie{} closed types)}. Constructing explicit models for the \gaurav{quantified} functions used to model Boogie's type system (satisfying\gaurav{,} \eg{}\gaurav{,} suitable inverse properties for the projection functions) is straightforward.
For the VC-level variable values, we can directly instantiate the \gaurav{corresponding} values in the initial Boogie program state.

VC-level functions representing those declared in the Boogie program are instantiated as (total) functions which, \emph{for input values of appropriate type} \gaurav{(the arguments and output are untyped values of sort $V$)}, are defined simply to return the same values as the corresponding function in our model. However, perhaps surprisingly, Boogie's VC embedding of functions logically
requires \gaurav{functions to return values of the specified return type even if the input values do not have the types specified by the function.}
\gout{requires properties of these functions even in other cases.}%
\gout{For example, for the \bcode{foo} function above, \emph{some} value of the type $\textit{List}^{\pi}_1(\typeofval{}{v})$ must be returned even for arguments which are not lists!}%
\gaurav{In such cases,} we define the \gaurav{instantiated} function to return some \gout{such}value \gaurav{of the specified type}, which is possible since in well-formed contexts every closed type has at least one value in our model.

\gout{Secondly}\gaurav{After our instantiation}, we need to prove the hypotheses of the VC's implication; in particular that all axioms (both those generated by the type system encoding and those coming from the program itself) are satisfied. The former are standard and simple to prove (given the work above), while the latter largely follow from the assumption \gout{on \emph{executions}}that each declared axiom must be satisfied in the initial state restricted to the constants. The only remaining challenge is to relate VC expressions with the evaluation of corresponding Boogie expressions; an issue which also arises (and is explained) below\gaurav{, where we show how to connect validity of the instantiated WP to the program}.

\subsection{Certifying the VC Phase}
Boogie's weakest precondition calculation is made size-efficient by the usage of explicit named constants for the weakest preconditions $\wpboogie{B}$ for each block $B$, which is defined in terms of the named constants for its successor blocks.
For example, in~\figref{fig:passification_example}, $\wpboogie{B_2''}$ is given by $i^{vc}_1 \neq 0 \Longrightarrow \wpboogie{B_3''} \wedge \wpboogie{B_4''}$.
Here $i^{vc}_1$ is the value that we instantiated for the variable \bcode{i1}.

We exploit this modular construction of the generated weakest precondition for the local and global block theorems.
We prove for each block $B$ with commands $cs$ the following local block lemma:
\begin{theorem}[VC Phase Local Block Lemma]
\\
If $\redcmdlist{A}{\Lambda}{\Gamma}{\Omega}{\mathit{cs}}{\normal{\mathit{ns}}}{s'}$ and $\wpboogie{B}$ holds, then $s' \neq \failure{}$ and if $s'$ is a normal state, then $\forall B_\mathit{suc} \in \mathit{successors}(B).\; \wpboogie{B_\mathit{suc}}$.
\end{theorem}
Once one has proved this lemma for all blocks in the CFG, combining them to obtain the corresponding global block theorems (via our usual reverse walk of the CFG) is straightforward.
The main challenge is in decomposing the proof for the local block lemma itself for a block $B$, for which we outline our approach next.

By this phase, the first command in $B$ must be either an $\assume{e}$ or an $\assert{e}$ command. In the former case, we rewrite $\wpboogie{B}$ into the form $e^{vc} \Longrightarrow H$, where $e^{vc}$ is the VC counterpart of $e$ and where $H$ corresponds to the weakest precondition of the remaining commands. This rewriting may involve undoing certain optimisations Boogie's implementation performed on the formula structure. Next, we need to prove that $e$ evaluates to $e^{vc}$ (see below). Hence, if $e$ evaluates to \btrue{} (the execution does not go to magic) then $H$ must be true, and we can continue inductively. The argument for $\assert{e}$ is similar but \gout{rewriting}\gaurav{where we rewrite} the VC to $e^{vc} \wedge H$ (\ie{}  $e^{vc}$ and $H$ must both hold); if $e$ evaluates to $e^{vc}$, we know that the execution does not fail.

%
Proving that $e$ evaluates to $e^{vc}$ arises in both cases and also in our previous discharging of VC hypotheses.
Note that\gaurav{, in contrast to $e$,} $e^{vc}$ is not a Boogie expression, but a shallowly embedded formula that includes the instantiations of quantified variables we constructed above. Showing this property works largely on syntax-driven rules that relate a Boogie expression with its VC counterpart, except for extra work due to mismatching function signatures  and optimisations that Boogie made either to the formula structure or via the type system encoding (\cf{}~\secref{sec:typesystem_desugaring}).
\gout{We handle some of these cases by making Isabelle prove that we can rewrite the formula back into the unoptimised standard form we require for our syntax-driven rules and in other cases we use Isabelle to prove the goal directly.}
\gaurav{We handle some of these cases by showing that we can rewrite the formula back into the unoptimised standard form we require for our syntax-driven rules and in other cases we directly work with the optimised form. Both cases are automated using Isabelle tactics.}

\bigskip
This concludes our discussion of the certification of Boogie's three key phases. Combining the three certificates yields an end-to-end proof that the validity of the generated verification conditions implies the correctness of the input program, that is, that the given verification run is sound.

\section{Implementation and Evaluation}\label{sec:evaluation}
In this section, we evaluate our certifying version of the Boogie verifier~\cite{artifact}, which produces Isabelle certificates proving the correctness of Boogie's pipeline for programs it verifies.

We have implemented our validation tool as a new C\# module compiled with Boogie.
We instrumented Boogie's codebase to call out to our module\gaurav{, which allows us to obtain}\gout{logging various} information that we \gaurav{can} use to validate the key phases, and extended parts of the codebase to extract information more easily.
Moreover, we disabled counter-example related VC features and the generation of VC axioms for any built-in types \gaurav{and operators} that we do not support.
We added or changed \gout{only 143}\gaurav{fewer than 250 non-empty, uncommented} lines of code across \gout{6}\gaurav{11} files in the existing Boogie implementation.

Given an input file verified by Boogie, our work produces an Isabelle certificate per procedure $p$ that certifies the correctness of \gout{its CFG}\gaurav{the corresponding CFG-to-DAG source CFG} as represented internally in Boogie.
\gaurav{The generation and checking of the certificate is fully automatic, without any user input. We use a combination of custom and built-in Isabelle tactics.}
In addition to the three key phases we describe in detail, our implementation also handles several smaller transformations made by Boogie, such as constant propagation.
Our tool currently supports the default options of Boogie (only) and does not support \gout{\eg{}}advanced source-level \emph{attributes} (\gout{usable \eg{}}\gaurav{for instance,} to selectively force procedures to be inlined).

\begin{table}[t]
\small
\centering
\caption{Selection of algorithmic examples with the lines of code (LOC), the number of procedures (\#P), the time it takes for Isabelle to check the certficate in seconds (the average of 5 runs on a Lenovo T480 with 32 GB, i7-8550U 1.8 GhZ, Ubuntu 18.04 on the Windows Subsystem for Linux), and the certificate size expressed as the number of non-empty lines of Isabelle.}
\begin{tabular}{l||c|c|c|c}
Name & LOC & \#P & Time [s] & Size \\
\hline
TuringFactorial & 29 & 1 & \gout{17.1}\gaurav{19.4}  & \gout{1994}\gaurav{1986} \\
Find & 27 & 2 & \gout{39.7}\gaurav{27.3} & \gout{2168}\gaurav{2100} \\
DivMod  & 69 & 2 & \gout{34.7}\gaurav{28.4} & \gout{4839}\gaurav{4753} \\
Summax~\cite{firstvercomp_experience} & 23 & 1& \gout{16.9}\gaurav{19.1} & \gout{1962}\gaurav{1953} \\
MaxOfArray~\cite{CF-iFM17} & 22 & 1 & \gout{17.1}\gaurav{19.9} & \gout{1949}\gaurav{1944} \\
SumOfArray~\cite{CF-iFM17} & 22 & 1 & \gout{17.6}\gaurav{18.7} & \gout{1539}\gaurav{1534} \\
Plateau~\cite{CF-iFM17} & 50 & 1 & \gout{20.8}\gaurav{22.9} & \gout{2024}\gaurav{2019}\\
WelfareCrook~\cite{CF-iFM17} & 52 & 1 & \gout{38.4}\gaurav{39.4} & \gout{2545}\gaurav{2528} \\
ArrayPartitioning~\cite{CF-iFM17} & 57 & 2 & \gout{38.0}\gaurav{27.6} & \gout{3606}\gaurav{3514} \\
DutchFlag~\cite{CF-iFM17} & 76 & 2 & \gout{66.4}\gaurav{52.8} & \gout{4124}\gaurav{3994}
\end{tabular}
\label{tbl:selected_examples}
\end{table}

We evaluated our work in two ways. Firstly, to evaluate the applicability of our certificate generation, we automatically collected all input files with at least one procedure from Boogie's \gout{own}test suite~\cite{boogierepo} which verify successfully and which either use no unsupported features or are easily desugared (by hand) into versions without them. This includes programs with procedure calls since Boogie simply desugars these in an early stage.  For programs employing \gout{unsupported}attributes, we checked whether the program still verifies \emph{without} attributes, and if so we also kept these. In total, this yields \gout{95}\gaurav{100} programs from Boogie's test suite. Secondly, we collected a corpus of ten Boogie programs which verify interesting algorithms with non-trivial specifications: three from Boogie's test suite and seven from the literature~\cite{CF-iFM17,firstvercomp_experience}. Where needed we manually desugared usages of Boogie maps (which \gout{our work does not}\gaurav{we do not} yet support) using type declarations, functions, and axioms.

Of the \gout{95}\gaurav{100} programs from Boogie's test suite, we successfully generate certificates in \gout{89}\gaurav{96} cases\gout{(93\%)}. The remaining \gout{6}\gaurav{4} cases involve special cases that we do not handle yet. \gout{For 4 of them, extending our work is straightforward: one special case is that Boogie's passification transforms a boolean assignment \bcode{x := e} to \assume{\bcode{xi} $\Leftrightarrow$ \bcode{e'}} instead of \assume{\bcode{xi = e'}}.
The other case includes a block with an edge directly to itself; this unusual case trips up our current implementation, which will be easily amended.}%
\gaurav{
For 2 of them, extending our work is straightforward: one special case includes a naming clash and the other case can be amended by using a more specific version of a helper lemma.
}
The remaining two fail because of our incomplete handling of function calls in the VC phase when combined with coercions between VC integers or booleans and their Boogie counterparts.
Handling this is more challenging but is not a fundamental issue.

For the corpus of 10 examples, \tabref{tbl:selected_examples} shows the generated certificate size and the time for Isabelle to check their validity.\footnote{The time to generate the certificate is not included, but is negligible here.}
The ratio of certificate size to code size ranges from \gout{40}\gaurav{41} to 89; this rather large ratio emphasises the substantial work in \gout{precisely and}formally validating the substantial work which Boogie's implementation performs. \gaurav{Optimisations to further reduce the ratio are possible.} The validation of certificates takes usually under one second per line of code. While these times are not short, they are acceptable since certificate generation needs to run only for \gout{a}\gout{final}(verified) \gaurav{release} version\gaurav{s} of the program in question. 

\section{Related Work}\label{sec:related_work}
Several works explore the validation of program verifiers.
Garchery et al.~\cite{garchery-why3} validate VC rewritings in the Why3 VC generator~\cite{why3}.
Unlike our work, they do not connect VCs with programs and do not handle the erasure of polymorphic types.
Strub et al.~\cite{self_certification_fstar} validate part of a previous version of the F* verifier~\cite{fstar16} by generating a certificate for the F* type checker itself, which type checks programs by generating VCs.
Like us, they assume the validity of the generated VC itself, but they do not consider program-to-program transformations such as ours.
Another approach \gout{to validate verification results}is taken by Aguirre~\cite{Aguirre16} who shows how one can map proofs of the VC back to correctness of an F* program.
\gout{,which could be used in conjunction with the proof-producing capability of modern SMT solvers;}%
They prove a once-and-for-all result, but the approach could be \gout{directly}lifted to \gaurav{a} validation \gaurav{approach using the proof-producing capability of SMT solvers~\cite{BdMF15}. Lifting the approach would require extending the work to handle classical instead of constructive VC proofs.}
\gout{However, the work has not been implemented, and makes various assumptions about the VC proof that are not guaranteed by SMT solvers such as the proof being constructive and being in a normal form.}

There is some work on proving VC generator implementations correct once and for all, although none of the proven tools are used in practice.
Homeier and Martin~\cite{Homeier1995} prove a VC generator correct in HOL for an executable \gout{programming}language and a simpler VC \gout{generation technique}\gaurav{phase} than Boogie's.
Herms et al.~\cite{herms_vstte12} prove a VC generator inspired by Why3 correct in Coq. However, some more-challenging aspects of Why3's VC transformation and polymorphic type system are not handled.
Vogels et al.~\cite{VogelsVCGProof2010} prove a toolchain for a Boogie-like language correct in Coq, including passification and VC phases.
However, the language is quite limited: without unstructured control flow, loops (\ie{} no need for a CFG-to-DAG phase), functions, or polymorphism (\ie{} no type encoding).
\gaurav{Verifiers other than VC generators, include the verified Verasco static analyzer~\cite{jourdan2015formally}, which supports a realistic subset of C, but whose performance is not yet on par with unverified, industrial  analyzers.}

\gaurav{Validation \gout{techniques have}\gaurav{has} also been \gout{used}\gaurav{explored} in other settings. Alkassar et al.~\cite{alkassar14} adjust graph algorithms to produce witnesses that can be then used by verified validators to check whether the result is correct.}
In the \gout{related}context of compiler correctness, many validation techniques express a per-run validator in Coq, prove it correct once-and-for-all~\cite{Tristan2008,compcertssa,Tristan2010}, and then extract executable code (the extraction must be trusted).
\gaurav{In the verified CompCert compiler~\cite{LeroyCompCert}, such validators have been used in combination with the once-and-for-all approach. Validators are used for phases that can be more easily validated than proved correct once and for all.}
One such \gout{work}\gaurav{example} related to our certification of the passification phase is the validation of the SSA phase~\cite{compcertssa}\gout{in CompCert}, dealing also with versioned variables in the target (but not with \assumeNoArg{} statements that prune executions).
In contrast to our work, they require an explicit notion of CFG domination and they do not use a global versioning scheme to efficiently check that two parts of the CFG constrain disjoint versions.
Our versioning idea is similar to a technique used for the validation of a dominator relation in a CFG~\cite{validation_dominators}, which assigns intervals to basic blocks (as opposed to assigning versions to variables) to efficiently determine whether a block dominates another one.
The validation of the Cogent compiler~\cite{CogentITP2016} follows a similar approach to ours in that it \gout{directly}generates proofs in Isabelle.

\section{Conclusion}\label{sec:conclusion}
We have presented \gout{and implemented}a novel verifier validation approach, and applied it successfully to three key phases of the Boogie verifier, providing formal underpinnings for both the language and its verifier for the first time.
Our work demonstrates that it is feasible to provide strong formal guarantees regarding the verification results of practical VC generators written in modern mainstream \gout{programming}languages.
\gout{
In the future, we plan to investigate the extension and
application of our overall validation approach to verification tools which map
verification problems concerning other languages and logics into intermediate
verification languages such as Boogie.
}

\gaurav{
In the future, we plan to extend our supported subset of Boogie, \asout{for instance,}\as{\eg{}} to include procedure calls and bitvectors.
Supporting \as{Boogie's potentially-impredicative} maps is the main \as{open} challenge\as{: maps can take other maps as input, potentially including themselves.}\asout{ since they can take maps as input which potentially include themselves.}
The challenge \as{with this feature} is to \as{still be able to} express a type \as{in Isabelle capturing}\asout{ that captures} all Boogie values \as{despite the}\asout{due to this} potentially-cyclic nature of map types. \as{In practice, however, this may not be required in full generality:}\asout{However, the full generality does not seem necessary in practice:} we have observed that Boogie front-ends \as{rarely}\asout{don't} use maps that contain maps of the same type as input. Therefore, we plan to extend our technique to support a suitably-expressive restricted form of Boogie maps.

}

\gaurav{
\subsubsection{Acknowledgements.}
We thank Alain Delaët–Tixeuil for his earlier work on this topic, Thibault Dardinier for improving our artifact, Martin Clochard for helpful discussions and the anonymous reviewers for their valuable comments.
This work was partially funded by the Swiss National Science Foundation (SNSF) under Grant No. 197065.
} 


\bibliographystyle{splncs04}
\bibliography{references}

\ifx\istr\undefined
\newpage
\end{document}
\fi
\newpage
\begin{appendix}
\section{A Formal Semantics for Boogie}\label{app:boogie_lang}
\subsection{The Boogie Language: Syntax}
\begin{figure}[t]
    \begin{align*}
    \mathit{bop} ::=&\; = | \neq | + | - | * | \;/\; | \;\mathit{mod}\; | \leq | < | \geq | > | \wedge | \vee | \Rightarrow | \Leftrightarrow  \quad\quad
    \mathit{uop} ::=\; - | \neg \\
    e ::=&\; x \mid \bfalse \mid \btrue \mid i \mid \binaryop{e_1}{\mathit{bop}}{e_2} \mid \unaryop{\mathit{uop}}{e} \mid \funcall{f}{\vec \tau}{\vec e} \mid \old{e} \mid \\
    &\; \bforall{x}{\tau}{e} \mid \bexists{x}{\tau}{e} \mid \bforallt{t}{e} \mid \bexistst{\tau}{e} \\
    \tau ::=&\; \mathit{Int} \mid \mathit{Bool} \mid \tcon{\mathit{C}}{\vec \tau} \mid t \quad
    c ::=\; \assume{e} \mid \assert{e} \mid \assign{x}{e} \mid \havoc{x}\\
    decls ::=&\; \axiom{e} \mid \fdecl{f}{\vec t}{\vec \tau}{\tau'} \mid \vdecl{x}{\tau} \mid
    \cdecl{x}{\tau} \\
    &\; \tcondecl{C}{n} \mid \procdecl{p}{e_\mathit{pre}}{e_\mathit{post}}{\vec p}{\vec r}{\vec l}{\mathit{G}}
    \end{align*}
    \caption{
        The syntax of our formalised Boogie subset, where $\tau$, $e$, $c$, $\mathit{decls}$ denote the types, expressions, basic commands, and top-level declarations respectively; control-flow is handled via CFGs over the basic commands. $\mathit{bop}$ and $\mathit{uop}$ denote binary and unary operations, respectively. We sometimes write \bcode{a != b} instead of \bcode{a} $\neq$ \bcode{b} and \bcode{!a} instead of $\neg$\bcode{a}. 
    }
    \label{fig:boogie_syntax_full}
\end{figure}
The types, expressions, basic commands, and top-level declarations in our Boogie subset are shown in~\figref{fig:boogie_syntax_full}.
\gaurav{
Top-level declarations include axioms, function declarations, global variable declarations, constant declarations, type constructor declarations, and  procedure declarations.
Functions can be polymorphic as indicated by the type parameters $\vec{t}$.
Type constructor declarations include the constructor name $C$ and the number of type parameters $n$.
Each procedure declaration includes a pre- and a postcondition ($e_\mathit{pre}$ and $e_\mathit{post}$), the parameter declarations ($\vec{p}$), the result variable declarations $\vec{r}$, and a body given by the local variable declarations ($\vec{l}$) and a CFG ($\mathit{G}$).
}

We support the primitive types $\mathit{Int}$ and $\mathit{Bool}$; types obtained via declared type constructors are \emph{uninterpreted types}; the sets of values such types denote are constrained only via Boogie axioms and \assumeNoArg{} commands.
\gaurav{Moreover, types can contain type variables (for instance, to specify polymorphic functions).}

Expressions include variables, Boolean and integer literals, unary and binary expressions.
We also support function calls $\funcall{f}{\vec \tau}{\vec e}$.
The arguments $\vec \tau$ to a function call $\funcall{f}{\vec \tau}{\vec e}$ instantiate any \emph{type} parameters and are inferred by the type-checker; in our formalization type parameters are always explicit.
Old-expressions $\old{e}$ evaluate the expression $e$ w.r.t.\ the current local data and the global data as it \emph{was} in the pre-state of the procedure execution.
The remaining expressions are value quantification ($\bforall{x}{\tau}{e}$/$\bexists{x}{\tau}{e}$), and type quantification ($\bforallt{t}{e}$/$\bexistst{t}{e}$).

The commands are given by assumptions, assertions, assignments and havoc commands.
Sequential composition is represented by basic blocks that contain a list of commands.

Boogie source programs contain richer expressions and commands that can be desugared straightforwardly into our subset. Examples include havocs of multiple variables and value quantification with multiple binders. Some, such as procedure calls, are already desugared by Boogie in pre-processing phases.

\subsection{Expression evaluation.}\label{app:red_expr}
The (big-step) rules for the expression evaluation are given in~\figref{fig:red_expr_basic} (basic expressions),~\figref{fig:red_expr_quantified} (quantified expressions), and~\figref{fig:red_expr_list} (lists of expressions).
The rule for variable lookup is defined in terms of the function $\lookupvar{(G,L)}{\mathit{gs}}{\mathit{ls}}{x}$, which returns $\mathit{ls}(x)$ if $x$ belongs to the local data (\ie{}  $x$ is recorded in the type declarations $L$ for the local data) and $\mathit{gs}(x)$ otherwise.\footnote{$(G,L)$ is a variable context where $G$ and $L$ are the type declarations for the global and local data.}
This models the fact that local variables shadow global variables.
In the rule for literals, $l_e$ and $l_v$ denote literal expressions and the corresponding literal values respectively.
The rules for value quantification are defined in terms of $\typeofval{\typs}{v}$, which maps a value $v$ to its type w.r.t. the type interpretation $\typs$.
\begin{figure}
    \[
    \begin{array}{ll}
    \Inf{\lookupvar{\Lambda}{gs}{ls}{x} = v}{\expreval{\typs}{\Lambda}{\Gamma}{\Omega}{x}{\normal{(\mathit{os},\mathit{gs},\mathit{ls})}}{v}} &
    \Inf{}{\expreval{\typs}{\Lambda}{\Gamma}{\Omega}{l_e}{\normal{\mathit{ns}}}{l_v}} \\[3em]
    \Inf{\expreval{\typs}{\Lambda}{\Gamma}{\Omega}{e_1}{\normal{\mathit{ns}}}{v_1}}.
    {\expreval{\typs}{\Lambda}{\Gamma}{\Omega}{e_2}{\normal{\mathit{ns}}}{v_2}}.
    {v_1 \;\overline{\mathit{bop}}\; v_2 = v}
    {\expreval{\typs}{\Lambda}{\Gamma}{\Omega}{e_1 \;\mathit{bop}\; e_2}{\normal{\mathit{ns}}}{v}} &
    \Inf{\expreval{\typs}{\Lambda}{\Gamma}{\Omega}{e}{\normal{\mathit{ns}}}{v'}}
    {\overline{uop}(v')=v}
    {\expreval{\typs}{\Lambda}{\Gamma}{\Omega}{\mathit{uop}(e)}{\normal{\mathit{ns}}}{v}}\\[3em]
    \Inf{\exprevallist{A}{\Lambda}{\Gamma}{\Omega}{\vec{e}}{\normal{\mathit{ns}}}{\vec{v'}}}.{\Gamma(f) = \overline{f}}{\overline{f}(\Omega(\vec{\tau}),\vec{v'})=v}{\expreval{\typs}{\Lambda}{\Gamma}{\Omega}{\funcall{f}{\vec{\tau}}{\vec{e}}}{\normal{\mathit{ns}}}{v}} &
    \Inf{\expreval{\typs}{\Lambda}{\Gamma}{\Omega}{e}{\normal{(\mathit{os},\mathit{os},\mathit{ls})}}{v}}{\expreval{\typs}{\Lambda}{\Gamma}{\Omega}{\old{e}}{\normal{(\mathit{os},\mathit{gs},\mathit{ls})}}{v}}
    \end{array}
    \]
    \caption{Expression evaluation for basic expressions}
    \label{fig:red_expr_basic}
    \end{figure}
\begin{figure}
\[
\begin{array}{l}
\textit{Value quantification} \\[1.5em]
\Inf{\forall w.\; \typeofval{\typs}{w} = \Omega(\tau) \Longrightarrow
\expreval{\typs}{\Lambda}{\Gamma}{\Omega}{e}{\normal{(\mathit{os},\mathit{gs},\mathit{ls}(x \mapsto w))}}{\btrue} }{\expreval{\typs}{\Lambda}{\Gamma}{\Omega}{\bforall{x}{\tau}{e}}{\normal{(\mathit{os},\mathit{gs},\mathit{ls})}}{\btrue}}\\[2em]
\Inf{\typeofval{\typs}{w} = \Omega(\tau)}
{\expreval{\typs}{\Lambda}{\Gamma}{\Omega}{e}{\normal{(\mathit{os},\mathit{gs},\mathit{ls}(x \mapsto w))}}{\bfalse} }{\expreval{\typs}{\Lambda}{\Gamma}{\Omega}{\bforall{x}{\tau}{e}}{\normal{(\mathit{os},\mathit{gs},\mathit{ls})}}{\bfalse}} \\[2em]
\Inf{\typeofval{\typs}{w} = \Omega(\tau)}
{\expreval{\typs}{\Lambda}{\Gamma}{\Omega}{e}{\normal{(\mathit{os},\mathit{gs},\mathit{ls}(x \mapsto w))}}{\btrue} }{\expreval{\typs}{\Lambda}{\Gamma}{\Omega}{\bexists{x}{\tau}{e}}{\normal{(\mathit{os},\mathit{gs},\mathit{ls})}}{\btrue}} \\[2em]
\Inf{\forall w.\; \typeofval{\typs}{w} = \Omega(\tau) \Longrightarrow
\expreval{\typs}{\Lambda}{\Gamma}{\Omega}{e}{\normal{(\mathit{os},\mathit{gs},\mathit{ls}(x \mapsto w))}}{\bfalse} }{\expreval{\typs}{\Lambda}{\Gamma}{\Omega}{\bexists{x}{\tau}{e}}{\normal{(\mathit{os},\mathit{gs},\mathit{ls})}}{\bfalse}} \\[2em]
\textit{Type quantification} \\[1.5em]
\Inf{\forall \tau.\; \isclosedty{\tau} \Longrightarrow \expreval{\typs}{\Lambda}{\Gamma}{\Omega(t \mapsto \tau)}{e}{\normal{\mathit{ns}}}{\btrue{}} }{\expreval{\typs}{\Lambda}{\Gamma}{\Omega}{\bforallt{t}{e}}{\normal{\mathit{ns}}}{\btrue{}}} \\[2em]
\Inf{\isclosedty{\tau}}{\expreval{\typs}{\Lambda}{\Gamma}{\Omega(t \mapsto \tau)}{e}{\normal{\mathit{ns}}}{\bfalse{}} }{\expreval{\typs}{\Lambda}{\Gamma}{\Omega}{\bforallt{t}{e}}{\normal{\mathit{ns}}}{\bfalse{}}} \\[2em]
\Inf{\isclosedty{\tau}}{\expreval{\typs}{\Lambda}{\Gamma}{\Omega(t \mapsto \tau)}{e}{\normal{\mathit{ns}}}{\btrue{}} }{\expreval{\typs}{\Lambda}{\Gamma}{\Omega}{\bexistst{t}{e}}{\normal{\mathit{ns}}}{\btrue{}}} \\[2em]
\Inf{\forall \tau.\; \isclosedty{\tau} \Longrightarrow \expreval{\typs}{\Lambda}{\Gamma}{\Omega(t \mapsto \tau)}{e}{\normal{\mathit{ns}}}{\bfalse{}} }{\expreval{\typs}{\Lambda}{\Gamma}{\Omega}{\bexistst{t}{e}}{\mathit{ns}}{\bfalse{}}}
\end{array}
\]
\caption{Expression evaluation for quantifiers.}
\label{fig:red_expr_quantified}
\end{figure}
\begin{figure}
\[
\begin{array}{ll}
\Inf{}{\exprevallist{A}{\Lambda}{\Gamma}{\Omega}{\textsf{nil}}{\normal{\mathit{ns}}}{\textsf{nil}}} &
\Inf{\expreval{A}{\Lambda}{\Gamma}{\Omega}{e}{\normal{\mathit{ns}}}{v}}.
{\exprevallist{A}{\Lambda}{\Gamma}{\Omega}{\mathit{es}}{\normal{\mathit{ns}}}{\mathit{vs}}}
{\exprevallist{A}{\Lambda}{\Gamma}{\Omega}{(e:\mathit{es})}{\normal{\mathit{ns}}}{(v:\mathit{vs})}}
\end{array}
\]
\caption{Expression evaluation for lists of expressions}
\label{fig:red_expr_list}
\end{figure}

\subsection{Command and CFG reduction.}\label{app:red_cmd_cfg}
The (big-step) rules for the reduction of commands and lists of commands is given in~\figref{fig:red_cmd} and~\figref{fig:red_cmd_list}.
Assignment reduces only if the value to be assigned has the right type, \ie{} assignment preserves well-typed states.
This restriction is not required for well-typed programs, but it nevertheless makes some reasoning easier.
The rules for assignment and havoc rely on \lookupty{\Lambda}{x} that maps variable $x$ to its declared type w.r.t. the variable context $\Lambda$ (if $x$ is defined), and on $\updatevar{\Lambda}{\mathit{ns}}{x}{v}$, which returns the state $\mathit{ns}$ where $x$ is updated to $v$ (ensuring that the local state is updated if $x$ is local and the global state otherwise).

The (small-step) rules for the CFG reduction are given in~\figref{fig:red_cfg}.
$\blockcmds{G}{b}$ gives the list of commands for block $b$ in CFG $G$.
\begin{figure}
    \[
    \begin{array}{ll}
    \Inf{\expreval{\typs}{\Lambda}{\Gamma}{\Omega}{e}{\normal{\mathit{ns}}}{\btrue{}}}{\redcmd{\typs}{\Lambda}{\Gamma}{\Omega}{\assert{e}}{\normal{\mathit{ns}}}{\normal{ns}}}&
    \Inf{\expreval{\typs}{\Lambda}{\Gamma}{\Omega}{e}{\normal{\mathit{ns}}}{\bfalse{}}}{\redcmd{\typs}{\Lambda}{\Gamma}{\Omega}{\assert{e}}{\normal{\mathit{ns}}}{\failure{}}} \\[2em]
    \Inf{\expreval{\typs}{\Lambda}{\Gamma}{\Omega}{e}{\normal{\mathit{ns}}}{\btrue{}}}{\redcmd{\typs}{\Lambda}{\Gamma}{\Omega}{\assume{e}}{\normal{\mathit{ns}}}{\normal{ns}}}&
    \Inf{\expreval{\typs}{\Lambda}{\Gamma}{\Omega}{e}{\normal{\mathit{ns}}}{\bfalse{}}}{\redcmd{\typs}{\Lambda}{\Gamma}{\Omega}{\assume{e}}{\normal{\mathit{ns}}}{\magic{}}} \\[2em]
    \Inf{\expreval{\typs}{\Lambda}{\Gamma}{\Omega}{e}{\normal{\mathit{ns}}}{v}}.
    {\lookupty{\Lambda}{x} = \tau}
    {\typeofval{\typs}{v} = \Omega(\tau)}.
    {ns' = \updatevar{\Lambda}{\mathit{ns}}{x}{v}}
    {\redcmd{\typs}{\Lambda}{\Gamma}{\Omega}{\assign{x}{e}}{\normal{\mathit{ns}}}{\normal{\mathit{ns'}}}}    &
    \Inf{\lookupty{\Lambda}{x} = \tau}
    {\typeofval{\typs}{v} = \Omega(\tau)}.
    {ns' = \updatevar{\Lambda}{\mathit{ns}}{x}{v}}
    {\redcmd{\typs}{\Lambda}{\Gamma}{\Omega}{\havoc{x}}{\normal{\mathit{ns}}}{\normal{\mathit{ns'}}}} \\[3em]
    \Inf{}{\redcmd{\typs}{\Lambda}{\Gamma}{\Omega}{c}{\magic{}}{\magic{}}} &
    \Inf{}{\redcmd{\typs}{\Lambda}{\Gamma}{\Omega}{c}{\failure{}}{\failure{}}}
    \end{array}
    \]
    \caption{Command reduction}
    \label{fig:red_cmd}
\end{figure}

\begin{figure}
    \[
    \begin{array}{ll}
    \Inf{}{\redcmdlist{\typs}{\Lambda}{\Gamma}{\Omega}{\textsf{nil}}{s}{s}}&
    \Inf{\redcmd{\typs}{\Lambda}{\Gamma}{\Omega}{c}{s}{s''}}.
    {\redcmdlist{\typs}{\Lambda}{\Gamma}{\Omega}{\mathit{cs}}{s''}{s'}}
    {\redcmdlist{\typs}{\Lambda}{\Gamma}{\Omega}{(c:\mathit{cs})}{\normal{\mathit{ns}}}{s'}}
    \end{array}
    \]
    \caption{Reduction for lists of commands}
    \label{fig:red_cmd_list}
\end{figure}

\begin{figure}
    \[
    \begin{array}{ll}
    \Inf{\blockcmds{G}{b} = \mathit{cs}}
    {b' \in \mathit{successors}(G,b)}.
    {\redcmdlist{\typs}{\Lambda}{\Gamma}{\Omega}{\mathit{cs}}{\normal{\mathit{ns}}}{\normal{\mathit{ns'}}}}
    {\redcfg{\typs}{\Lambda}{\Gamma}{\Omega}{G}{(\inl{b}, \normal{\mathit{ns}})}{(\inl{b'}, \normal{\mathit{ns'}})}} \\[2em]
    \Inf{\blockcmds{G}{b} = \mathit{cs}}
    {\mathit{successors}(G,b) = \emptyset}.
    {\redcmdlist{\typs}{\Lambda}{\Gamma}{\Omega}{\mathit{cs}}{\normal{\mathit{ns}}}{\normal{\mathit{ns'}}}}
    {\redcfg{\typs}{\Lambda}{\Gamma}{\Omega}{G}{(\inl{b}, \normal{\mathit{ns}})}{(\inr{()}, \normal{\mathit{ns'}})}}\\[2em]
    \Inf{\blockcmds{G}{b} = \mathit{cs}}.
    {\redcmdlist{\typs}{\Lambda}{\Gamma}{\Omega}{\mathit{cs}}{\normal{\mathit{ns}}}{\magic{}}}
    {\redcfg{\typs}{\Lambda}{\Gamma}{\Omega}{G}{(\inl{()}, \normal{\mathit{ns}})}{(\inr{b'}, \magic{})}} \\[2em]
    \Inf{\blockcmds{G}{b} = \mathit{cs}}.
    {\redcmdlist{\typs}{\Lambda}{\Gamma}{\Omega}{\mathit{cs}}{\normal{\mathit{ns}}}{\failure{}}}
    {\redcfg{\typs}{\Lambda}{\Gamma}{\Omega}{G}{(\inl{()}, \normal{\mathit{ns}})}{(\inr{b'}, \failure{})}}
    \end{array}
    \]
    \caption{CFG reduction}
    \label{fig:red_cfg}
\end{figure}

\section{The Phases For the Running Example}\label{app:running_example}
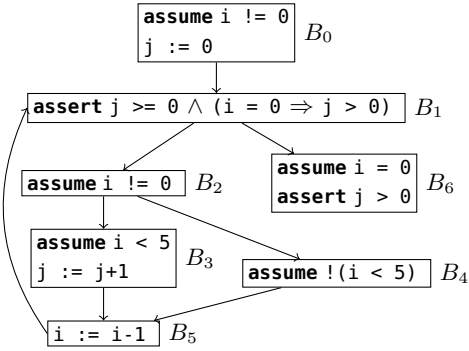
\begin{figure}
    \centering
\begin{tikzpicture}[baseline=(current bounding box.north)]
    \tikzset{vertex/.style = {shape=rectangle, draw, inner sep=2}}
    \tikzset{edge/.style = {->,> = latex'}}
    
    \node[align=left, vertex, label=right:{$B_0$}](nbefore) at (0,5) {\assume{\bcode{i != 0}}\\
    \bcode{j := 0}};
    \node[align=left, vertex, label=right:{$B_1$}](nhead) at (0,4) {
        \assert{\bcode{j >= 0} $\wedge$ \bcode{(i = 0} $\Rightarrow$ \bcode{j > 0)}}
    };
    \node[align=left, vertex, label=right:{$B_2$}](nbody) at (-1.5,3) {
        \bcodestandard{\assume{\bcode{i != 0}}}
    };
    \node[align=left, vertex, label=right:{$B_3$}](nleft) at (-1.5,2) {
        \assume{\bcode{i < 5}} \\
        \bcode{j := j+1}
    };
    \node[align=left, vertex, label=right:{$B_4$}](nright) at (1.6,1.8) {
        \assume{\bcode{!(i < 5)}}
    };
    \node[align=left, vertex, label=right:{$B_5$}](njoin) at (-1.5,1) {
        \bcode{i := i-1}
    };
    \node[align=left, vertex, label=right:{$B_6$}](nafter) at (1.7,3) {
        \assume{\bcode{i = 0}} \\
        \assert{\bcode{j > 0}}
    };

    \path
    (nbefore) [->] edge node [above] {} (nhead)
    (nhead) [->] edge [above] node [above] {} (nbody)
    (nhead) [->] edge node [above] {} (nafter)
    (nbody) [->] edge node [above] {} (nleft)
    (nbody) [->] edge node [above] {} (nright)
    (nleft) [->] edge node [above] {} (njoin)
    (nright) [->] edge node [above] {} (njoin)
    (njoin.west) [->] edge [bend left=30] node [above] {} (nhead.west);
\end{tikzpicture}
\caption{CFG representation of the running example}
\label{fig:running_example_cfg_full_app}
\end{figure}

\begin{figure}
    \centering
\begin{tikzpicture}[baseline=(current bounding box.north)]
    \tikzset{vertex/.style = {shape=rectangle, draw, inner sep=2}}
    \tikzset{edge/.style = {->,> = latex'}}
    
    \node[align=left, vertex, label=right:{$B'_0$}](nbefore) at (0,5.3) {\assume{\bcode{i != 0}}\\
    \bcode{j := 0} \\
    {\assert{\bcode{j >= 0} $\wedge$ \bcode{(i = 0} $\Rightarrow$ \bcode{j > 0)}}}};
    \node[align=left, vertex, label=right:{$B'_1$}](nhead) at (0,4) {
        {\havoc{\bcode{i,j}}} \\
        {\assume{\bcode{j >= 0} $\wedge$ \bcode{(i = 0} $\Rightarrow$ \bcode{j > 0)}}}
    };
    \node[align=left, vertex, label=right:{$B'_2$}](nbody) at (-1.5,3) {
        \bcodestandard{\assume{\bcode{i != 0}}}
    };
    \node[align=left, vertex, label=right:{$B'_3$}](nleft) at (-1.5,2) {
        \assume{\bcode{i < 5}} \\
        \bcode{j := j+1}
    };
    \node[align=left, vertex, label=right:{$B'_4$}](nright) at (1.6,1.8) {
        \assume{\bcode{!(i < 5)}}
    };
    \node[align=left, vertex, label=right:{$B'_5$}](njoin) at (-1.5,0.6) {
        \bcode{i := i-1} \\
        {\assert{\bcode{j >= 0} $\wedge$ \bcode{(i = 0} $\Rightarrow$ \bcode{j > 0)}}} \\
        {\assume{\bfalse{}}}
    };
    \node[align=left, vertex, label=right:{$B'_6$}](nafter) at (1.7,3) {
        \assume{\bcode{i == 0}} \\
        \assert{\bcode{j > 0}}
    };

    \path
    (nbefore) [->] edge node [above] {} (nhead)
    (nhead) [->] edge [above] node [above] {} (nbody)
    (nhead) [->] edge node [above] {} (nafter)
    (nbody) [->] edge node [above] {} (nleft)
    (nbody) [->] edge node [above] {} (nright)
    (nleft) [->] edge node [above] {} (njoin)
    (nright) [->] edge node [above] {} (njoin);
\end{tikzpicture}
\caption{CFG representation of the running example after the CFG-to-DAG phase}
\label{fig:running_example_dag_full_app}
\end{figure}
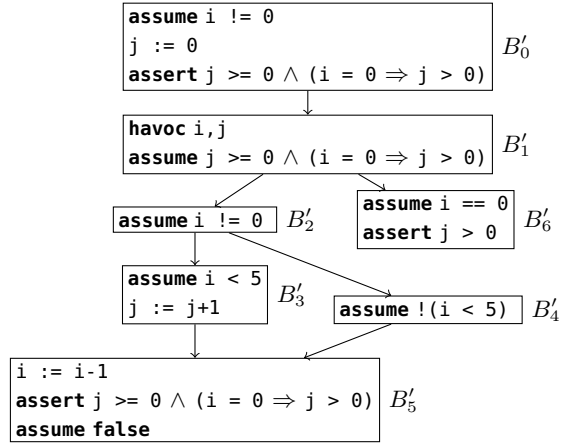

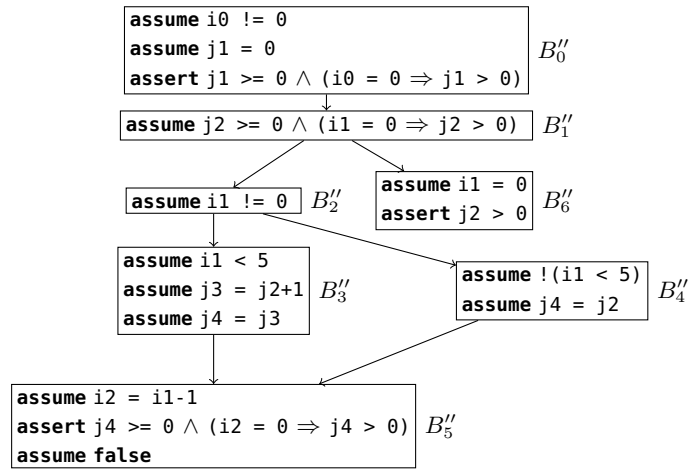
\begin{figure}
    \centering
\begin{tikzpicture}[baseline=(current bounding box.north)]
    \tikzset{vertex/.style = {shape=rectangle, draw, inner sep=2}}
    \tikzset{edge/.style = {->,> = latex'}}
    
    \node[align=left, vertex, label=right:{$B''_0$}](nbefore) at (0,5) {\assume{\bcode{i0 != 0}}\\
    \assume{\bcode{j1 = 0}} \\
    {\assert{\bcode{j1 >= 0} $\wedge$ \bcode{(i0 = 0} $\Rightarrow$ \bcode{j1 > 0)}}}};
    \node[align=left, vertex, label=right:{$B''_1$}](nhead) at (0,4) {
        {\assume{\bcode{j2 >= 0} $\wedge$ \bcode{(i1 = 0} $\Rightarrow$ \bcode{j2 > 0)}}}
    };
    \node[align=left, vertex, label=right:{$B''_2$}](nbody) at (-1.5,3) {
        \bcodestandard{\assume{\bcode{i1 != 0}}}
    };
    \node[align=left, vertex, label=right:{$B''_3$}](nleft) at (-1.5,1.8) {
        \assume{\bcode{i1 < 5}} \\
        \assume{\bcode{j3 = j2+1}} \\
        \assume{\bcode{j4 = j3}}
    };
    \node[align=left, vertex, label=right:{$B''_4$}](nright) at (3,1.8) {
        \assume{\bcode{!(i1 < 5)}} \\
        \assume{\bcode{j4 = j2}}
    };
    \node[align=left, vertex, label=right:{$B''_5$}](njoin) at (-1.5,0) {
        \assume{\bcode{i2 = i1-1}} \\
        \assert{\bcode{j4 >= 0} $\wedge$ \bcode{(i2 = 0} $\Rightarrow$ \bcode{j4 > 0)}} \\
        {\assume{\bfalse{}}}
    };
    \node[align=left, vertex, label=right:{$B''_6$}](nafter) at (1.7,3) {
        \assume{\bcode{i1 = 0}} \\
        \assert{\bcode{j2 > 0}}
    };

    \path
    (nbefore) [->] edge node [above] {} (nhead)
    (nhead) [->] edge [above] node [above] {} (nbody)
    (nhead) [->] edge node [above] {} (nafter)
    (nbody) [->] edge node [above] {} (nleft)
    (nbody) [->] edge node [above] {} (nright)
    (nleft) [->] edge node [above] {} (njoin)
    (nright) [->] edge node [above] {} (njoin);
\end{tikzpicture}
\caption{CFG representation of the running example after the passification phase}
\label{fig:running_example_passive_full_app}
\end{figure}

For our running example in~\figref{fig:running_example_ast_cfg}, the full CFG is shown in~\figref{fig:running_example_cfg_full_app}. 
The full CFG after the CFG-to-DAG phase is shown in~\figref{fig:running_example_dag_full_app}.
Finally, the full CFG after the passification phase is shown in~\figref{fig:running_example_passive_full_app}.
In practice, Boogie applies a constant propagation transformation as part of the passifcation phase.
Moreover, multiple empty blocks are added as well during the three phases.
We ignore both these points here for the sake of presentation, but we handle them 
in our validation tool.

\end{appendix}
\end{document}